\documentclass[nonacm,sigplan]{acmart}

\usepackage[utf8]{inputenc} 
\usepackage[T1]{fontenc}    
\usepackage{hyperref}       
\usepackage{url}            
\usepackage{booktabs}       
\usepackage{nicefrac}       
\usepackage{microtype}      
\usepackage{xcolor}         
\usepackage{subcaption}
\usepackage{caption}
\usepackage{wrapfig}
\usepackage{algorithm}
\usepackage{algpseudocode}
\usepackage{makecell}
\usepackage{multirow} 
\usepackage{pifont}
\newcommand{\cmark}{\ding{51}}%
\newcommand{\xmark}{\ding{55}}%

\usepackage{amsmath} 

\usepackage{graphicx} 
\usepackage{enumitem} 

\begin{document}

\title{xDiT: an Inference Engine for Diffusion Transformers (DiTs) with Massive Parallelism}

\author{Jiarui Fang\footnotemark[1]\footnotemark[2]}
\affiliation{
\institution{Tencent}
\country{China}
}
\email{jiaruifang@tencent.com}

\author{Jinzhe Pan\footnotemark[1]}
\affiliation{
\institution{Tencent \& Huazhong University of Science and Technology}
\country{China}
}
\email{eigensystem@hust.edu.cn}

\author{Xibo Sun}
\affiliation{
\institution{Tencent}
\country{China}
}
\email{xibosun@tencent.com}

\author{Aoyu Li}
\affiliation{
\institution{Tencent}
\country{China}
}
\email{aoyuli@tencent.com}

\author{Jiannan Wang}
\affiliation{
\institution{Tencent \& The University of Hong Kong}
\country{China}
}
\email{wchiennan@gmail.com}

\begin{abstract}

Diffusion models are pivotal for generating high-quality images and videos. Inspired by the success of OpenAI's Sora~\cite{sora2024}, the backbone of diffusion models is evolving from U-Net to Transformer, known as Diffusion Transformers (DiTs). However, generating high-quality content necessitates longer sequence lengths, exponentially increasing the computation required for the attention mechanism, and escalating DiTs inference latency. Parallel inference is essential for real-time DiTs deployments, but relying on a single parallel method is impractical due to poor scalability at large scales.

This paper introduces xDiT, a comprehensive parallel inference engine for DiTs. After thoroughly investigating existing DiTs parallel approaches, xDiT chooses Sequence Parallel (SP) and PipeFusion, a novel Patch-level Pipeline Parallel method, as intra-image parallel strategies, alongside CFG parallel for inter-image parallelism. xDiT can flexibly combine these parallel approaches in a hybrid manner, offering a robust and scalable solution.
Experimental results on two 8$\times$L40 GPUs (PCIe) nodes interconnected by Ethernet and an 8$\times$A100 (NVLink) node showcase xDiT's exceptional scalability across five state-of-the-art DiTs. Notably, we are the first to demonstrate DiTs scalability on Ethernet-connected GPU clusters.
xDiT is available at \url{https://github.com/xdit-project/xDiT}.

\end{abstract}

\maketitle 

\pagestyle{plain} 

\footnotetext[1]{* Jiarui Fang and Jinzhe Pan contributed equally to this work.}
\footnotetext[2]{$\dag$ Jiarui Fang is the corresponding author}

\section{Introduction}
In recent years, diffusion models have emerged as a groundbreaking technique in image~\cite{blackforestlabs2023announcement, esser2024scaling, chen2024pixart, peebles2023scalable} and video generation~\cite{sora2024, meta2024moviegen}. These models create images or videos through a multi-step denoising process, leveraging a neural network model at each step. We are witnessing a transformative shift in the architecture of the denoising networks. While traditionally dominated by U-Net~\cite{ronneberger2015u} architectures, these networks are now evolving into Diffusion Transformers (DiTs)~\cite{peebles2023scalable}, renowned for their superior model capacity and scalability. 

Images are encoded into a token sequence as input to DiTs~\cite{rombach2022high}.
The attention mechanisms of DiTs require mutual computation between tokens, leading to computation that scales quadratically with sequence length. 
Unlike Large Language Models (LLMs), which can generate tokens on the fly, DiTs must complete all steps to produce the final results. 
This poses significant challenges for real-time deployment.
For example, a Sora-like video generation API~\cite{mochi1preview_guide, kuaishou2024} takes over 4 minutes to generate a fewer seconds video. Given the immense computational demands, real-time DiTs deployment inevitably necessitates parallelism across multiple computing devices.

However, there is still a lack of effective methodology to scale DiTs to large scales. 
Although, several sequence parallelism (SP) methods~\cite{jacobs2023deepspeed, liu2023ring, fang2024unified, gu2024loongtrain} have been developed to scale long-sequence DiTs inference. Additionally, some approaches leverage \textit{input temporal redundancy}~\cite{so2024frdiff, ma2024deepcache, ma2024learning, yuan2024ditfastattn, zhao2024real, meta2024moviegen}, indicating a high degree of similarity in both inputs and activations across successive diffusion time steps, to design asynchronous sequence parallelism~\cite{li2023distrifusion} and patch-level sequence parallelism~\cite{wang2024pipefusion}. However, using these methods in isolation fails to adapt to the underlying heterogeneous interconnects of computational devices.
For instance, methods using collective communication are only suitable for high-bandwidth interconnects like NVLink, while methods using P2P (Peer-to-Peer) are more suitable for PCIe or Ethernet but do not have an advantage on NVLink.

Additionally, the diversity of DiTs model architectures presents challenges for parallel implementation. Unlike LLMs, which generally have a more uniform architecture, DiTs exhibit greater variability. For instance, the method of injecting user input instructions, the connection methods of DiTs blocks, and the operators' layouts in Transformer blocks can vary significantly. This diversity means that directly applying methods designed for LLMs, such as TP and SP, may not be immediately suitable for DiTs.

This paper introduces xDiT, a parallel inference system designed for DiTs. 
We argue that DiTs inference is analogous to the training of LLMs in that a single parallel is unlikely to scale effectively across various model architectures and network hardware, especially for those heterogeneous low-bandwidth networks. Therefore, a hybrid approach combining multiple parallel methods is necessary to achieve optimal communication efficiency.

Our contributions are as follows:

\begin{itemize}
\item We systematically investigate existing parallel methods for DiTs (including two types of sequence parallelism, DistriFusion, tensor parallelism, and PipeFusion), 
\item We selected PipeFusion and sequence parallelism for intra-image parallelism and CFG parallelism for inter-image parallelism. We designed a method to correctly hybrid these parallel approaches, thereby scaling DiTs inference to a very large scale.
\item Using the aforementioned methods, we built a system called xDiT. On 16$\times$L40 PCIe and 8$\times$A100 NVLink devices, we studied the scalability of five image and video generation DiTs (Pixart, Stable-Diffusion3, Flux.1, HunyuanDiT, CogVideoX). xDiT is the first system to successfully scale DiTs inference to 16 GPUs.
\end{itemize}

\section{Background \& Related Works}

\textbf{Diffusion Models}: Diffusion models utilize a noise-prediction deep neural network (DNN) denoted by $\epsilon_{\theta}$ to generate a high-quality image.
The process starts from pure Gaussian noise $x_T \sim \mathcal{N}(0, I)$ and involves numerous iterative denoising steps to produce the final meaningful image $x_0$, with $T$ representing the total number of diffusion time steps. 
At each diffusion time step $t$, given the noisy image $x_t$, the model $\epsilon_{\theta}$ takes $x_t$, $t$, and an additional condition $c$ (e.g., text, image) as inputs to predict the corresponding noise $\epsilon_t$ within $x_t$. 
At each denoising step, the previous image $x_{t-1}$ can be obtained from the following equation:
\begin{equation}
    x_{t-1} = \text{Update}(x_t, t, \epsilon_t), \quad \epsilon_t = \epsilon_{\theta}(x_t, t, c).
\end{equation}

In this context, \text{Update} denotes a function that is specific to the sampler, i.e. DDIM~\cite{song2020denoising} and DPM~\cite{lu2022dpm}, generally involves operations such as element-wise operations.
After multiple steps, we decode $x_0$ from the Latent Space to the Pixel Space using a Variational Autoencoder (VAE)~\cite{kingma2013auto}.
Consequently, the predominant contributor to diffusion model inference latency is attributed to the forward propagation through the model $\epsilon_{\theta}$.




\textbf{Diffusion Transformers (DiTs)}: The architecture of diffusion model $\epsilon_{\theta}$ is undergoing a pivotal transition from U-Net~\cite{ronneberger2015u} to Diffusion Transformers (DiTs)~\cite{meta2024moviegen, ma2024latte, chen2023pixart, bao2023all, li2024hunyuan}, driven by the scaling law demonstrating increased model parameters and training data with enhanced model performance. 
Unlike U-Nets, which apply convolutional layers capturing spatial hierarchies, DiTs encodes the input into a token sequence in the latent space and leverage the transformer's self-attention mechanism to model relationships within and across these patches.

\begin{figure}[t]
\centering
\includegraphics[width=0.4\textwidth]{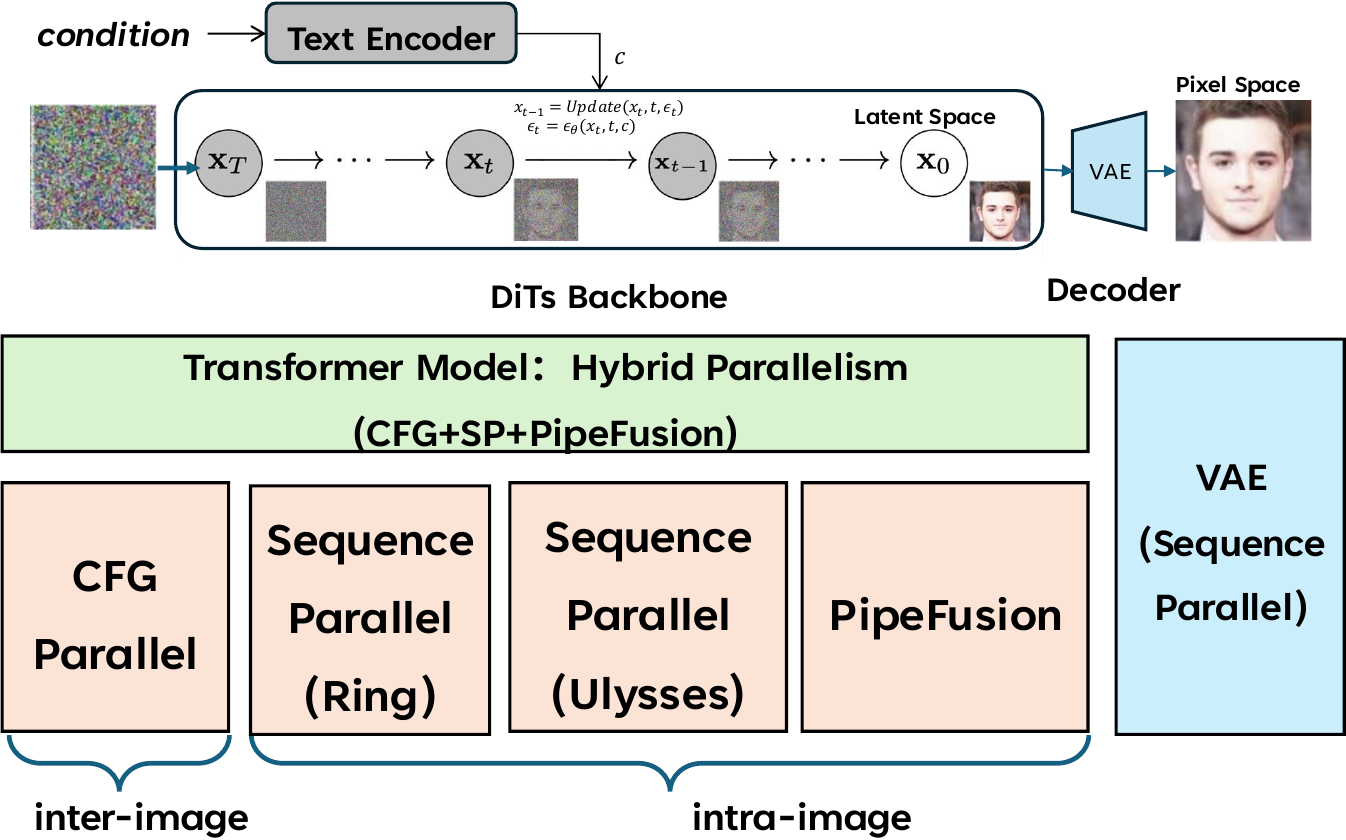}
\caption{The overview of xDiT system.}
\label{fig:xDiTovewview}
\end{figure}

As shown in the above of Figure~\ref{fig:xDiTovewview}, in DiTs, the input noisy latent representation is embedded into tokens and fed into a series of DiT blocks.
DiT blocks generally incorporate Multi-Head Self-Attention, Layer Norm, and Pointwise Feedforward Networks.
Although the earliest DiTs model architectures~\cite{peebles2023scalable} was quite similar to standard transformers~\cite{vaswani2017attention}, numerous improvements and variants have been introduced recently, leading to a highly diverse and non-uniform model architecture landscape.
For instance, the incorporation of conditioning can be achieved through various methods such as adaptive layer norm~\cite{peebles2023scalable}, cross-attention~\cite{chen2023pixart}, and extra input tokens~\cite{esser2024scaling}. 
As diffusion models tackle higher-resolution images and longer visual sequences, they impose a quadratic computational burden on inference.

\textbf{Input Temporal Redundancy:}
The diffusion model involves iterative noise prediction from input images or videos. Recent studies have emphasized \textit{input temporal redundancy}, highlighting the similarity in both inputs and activations across successive diffusion timesteps~\cite{so2024frdiff, ma2024deepcache}. Recent work ~\cite{ma2024learning} further explores the distribution of this similarity across various layers and timesteps. Leveraging this redundancy, some research caches activation values and reuses them in subsequent timesteps to reduce computation. For instance, in the U-Net architecture, DeepCache updates low-level features while reusing high-level ones from cache~\cite{ma2024deepcache}. Similarly, \textsc{Tgate} caches cross-attention outputs once they converge during the diffusion process~\cite{ma2024deepcache}. In contrast, for DiT models, $\Delta$-DiT caches rear DiT blocks in early sampling stages and front DiT blocks in later stages~\cite{chen2024delta}. PAB~\cite{zhao2024real} employs a pyramid-style broadcasting approach to mitigate temporal redundancy using a U-shaped attention pattern. Lastly, DiTFastAttn~\cite{yuan2024ditfastattn} identifies three types of redundancies—spatial, temporal, and conditional—and introduces an attention compression method to accelerate generation.

\textbf{Parallel Inference for Diffusion Model}:
Given the similar transformer architecture, tensor parallelism~\cite{shoeybi2019megatron} and sequence parallelism~\cite{jacobs2023deepspeed, liu2023ring, fang2024unified, gu2024loongtrain}, commonly used for efficient inference in LLMs, can be adapted for DiTs.
Tensor parallelism (TP) partitions model parameters across multiple devices, enabling parallel computation and reducing memory demands on individual devices. However, it requires AllReduce operations for the outputs of both the Attention and Feedforward Network modules, leading to communication overhead proportional to the sequence length. This overhead becomes significant for DiTs with very long sequences.
In contrast, sequence parallelism (SP) partitions the input image across multiple devices, using All2All or P2P to communicate Attention input and output tensors. SP offers better communication efficiency than TP but requires each device to store the entire model parameters, which can be memory-intensive.
DistriFusion~\cite{li2023distrifusion} exploits Input Temporal Redundancy to design an asynchronous sequence parallelism method for U-Net-based models, utilizing stable activations.
PipeFusion~\cite{wang2024pipefusion}, proposed by us, is a Patch-level Pipeline Parallelism. It leverages Input Temporal Redundancy to apply TeraPipe~\cite{li2021terapipe} to full attention DiTs inference, achieving lower communication overhead and model parameter memory usage compared to SP and DistriFusion.

\section{Challenges in Parallel DiTs Inference}

For a long time, inference of Diffusion Models with a U-Net backbone has been conducted on a single GPU due to limited computational requirements. 
However, in DiTs, the computation of the attention mechanism scales quadratically with the sequence length. 
With the scaling law~\cite{peebles2023scalable, liang2024scaling} driving improvements in image and video generation quality through increased model size and sequence length, it leads to a cubic growth in computational demand during deployment. 

In high-quality image and video generation tasks, the sequence length of the input to transformers can exceed 1 million tokens. 
The current leading open-source image generation model, Flux.1, generates images with a resolution of 1024px (1024$\times$1024), requiring a sequence length of 262 thousand tokens; For 4096px resolution images, the input sequence includes 4.2 million tokens. 
The leading open-source video generation model, CogVideoX, generates a 6-second video at 480x720 resolution with a sequence containing 17K tokens. If used to generate a one-minute 4K (3840$\times$2160) video, the sequence length exceeds 4 million.

Despite both applying Transformers, DiTs inference differs fundamentally from LLM inference.
DiTs forward a denoise network in multiple steps, with the final denoised result adopted as the meaningful output. 
In contrast, LLMs use autoregressive models, consisting of a Prefill phase for processing prompts and generating the first token, followed by a Decoding phase where the remaining tokens are generated sequentially. 
Both DiTs and the Prefill phase of LLMs are compute-bound, with DiTs particularly notable for handling extremely long input sequences, whereas the LLM Decode phase is memory-bound.
Therefore, applying LLM inference systems~\cite{vllm_project_2023} directly to DiTs is infeasible.
Parallel inference of DiTs presents significant challenges from two aspects:

\textit{\textbf{Challenges on Communication and Memory Cost}}:

\begin{figure*}[t]
\centering
\includegraphics[width=0.7\textwidth]{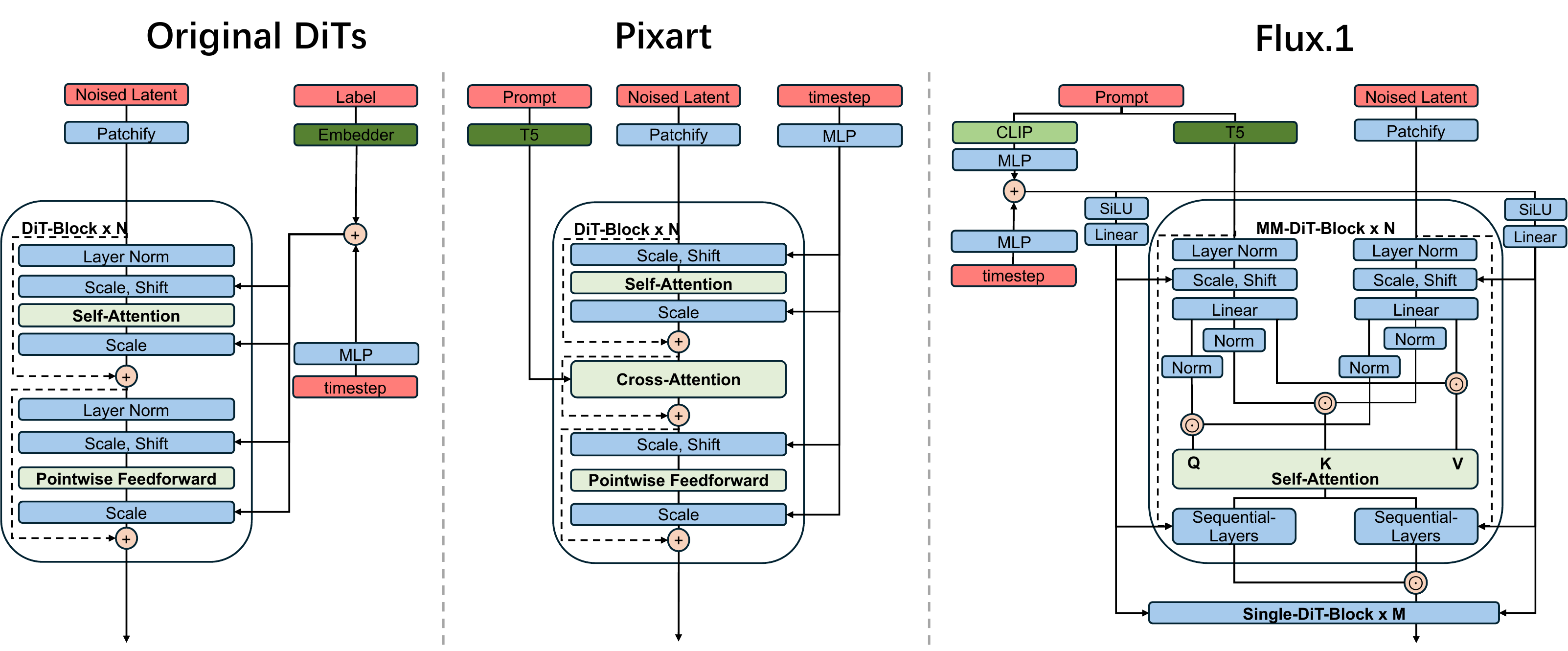}
\caption{The differences in DiTs model architectures. For the condition injection method, the original DiTs applies AdaLM-ZeRO; Pixart applies Cross-Attention; Flux.1 applies In-Context Condition.
The operator layouts inside a DiT block are also significantly different.}
\label{fig:dit_archs}
\end{figure*}

The current research on parallelizing DiTs primarily focuses on the scalability of individual parallel approaches. However, empirical evidence highlights the inadequacy of a single parallel method in managing the complexities of network connections between computing devices. Consequently, there is a notable gap in the literature regarding how to scale DiTs inference to extremely large scales, such as multiple GPU nodes, while accommodating both high-end networks with RDMA+NVLink and low-end networks with Ethernet+PCIe. The communication methods of a single parallel approach are ill-suited to handle the varying hardware network scenarios. Therefore, DiTs necessitates a hybrid parallel method to effectively adapt to different hardware networks.

Additionally, long sequences and large models impose significant memory demands. 
The current state-of-the-art open-source image generation model~\cite{blackforestlabs2023announcement} reached 12 billion parameters and is expected to grow further~\cite{fei2024scaling}. Without tensor parallelism, the designed parallel methods cannot effectively distribute parameters across multiple devices, such as sequence parallelism, which will lead to out-of-memory (OOM) issues. 
The VAE component of DiTs for image decoding also requires substantial activations and operator temporal memory.

\textit{\textbf{Challenges on Diverse DiTs model architectures}}:
As shown in Figure~\ref{fig:dit_archs}, starting from the original DiTs~\cite{peebles2023scalable}, the model architecture of DiTs has rapidly evolved, giving rise to many variants, each with its differences:

Firstly, the method of injecting condition information is not uniform. The original DiT~\cite{peebles2023scalable} proposes three condition injection methods: AdaLN-Zero, Cross-Attention, and In-Context Condition, which use the conditioning tensor to affect the Layer Norm, perform Cross-Attention computation, and concatenate them on sequence dimension, respectively. The first two methods only require splitting the image computation load, while the third method needs to consider splitting both the image and caption Transformers computation.

Secondly, despite incorporating both Self-Attention and Feedforward Network, there are significant differences in the implementation of DiT Blocks. For instance, the MM-DiT (Multimodal-DiT) blocks, adopted by Flux.1 and SD3, handle the condition latents and image latents equally. 
Prior to the Self-Attention module, QKV projections are conducted on both text and image latents, which are then concatenated on the sequence dimension. 
Due to this structure, designing a tensor parallelism strategy for MM-DiT becomes challenging. As it needs a sophisticated method to distribute and synchronize computations involving both types of latents across devices while maintaining the model's effectiveness.

Thirdly, the connection method between DiT blocks also varies. Compared to the initial linear connection, U-ViT~\cite{bao2023all} and HunyuanDiT~\cite{li2024hunyuan} designs skip-connected DiT blocks, which is similar to the U-Net's U-shaped topology, which presents the challenge for efficient communication pattern for pipeline parallels.

\section{xDiT System}
In this section, we present xDiT, a system designed for parallel inference of DiTs. 
As illustrated in Figure~\ref{fig:xDiTovewview}, DiTs inference is decomposed into three parts: Text-Encoder, Transformers Model, and VAE.
The majority of the computation is concentrated within the transformers model, which is the focal point of our paper. 
To address this, xDiT leverages two parallel paradigms: Intra-image Parallel (Sec.~\ref{sec:intra-image-parallel}) and Inter-image Parallel (Sec.~\ref{sec:inter-image-parallel}). 
The Intra-image paradigm employs multiple devices to process a single image input, relying on Sequence Parallelism (Sec.~\ref{sec:sp}) and an innovative parallel method named PipeFusion first proposed by us (Sec.~\ref{sec:pipefusion}). 
On the other hand, the Inter-image paradigm, which is used in conjunction with CFG (Classifier-Free Guidance), involves the parallel generation of images across devices. 
These four parallel methods can be combined in any arbitrary hybrid configuration. 
Furthermore, to prevent Out-of-Memory (OOM) issue in the VAE, we have devised a Patch Parallel method for it (Sec.~\ref{sec:vae}).

\subsection{Intra-Image Parallelism for Transformer Model}
\label{sec:intra-image-parallel}
This section will present how to leverage multiple computing devices to parallelize the computation of a single image in the DiTs backbone. 
We first adapt SP to DiT Block for DiTs models applying In-Context Conditioning.
The we review the Patch-level Pipeline Parallelism (PipeFusion).
Third, we compare the communication and memory efficiency of existing parallel approaches and our proposed methods.
Last but not least, we present an approach to correctly hybridize the parallel of SP and PipeFusion.

\subsubsection{Sequence Parallelism}
\label{sec:sp}
In DiTs, 2D latent images can be flattened and interpreted as a sequence of visual tokens, which then serve as inputs to the transformer blocks.
Sequence parallelism splits inputs along the sequence dimension by partitioning the input image into non-overlapping patches. 
Then each device computes the outputs for its local patches.
The best practices for sequence parallelism are DeepSpeed-Ulysses~\cite{jacobs2023deepspeed} (SP-Ulysses), Ring Atention~\cite{liu2023ring} (SP-Ring).
As shown in Figure~\ref{fig:twosp}, SP-Ulysses employs All2All communications to transform the partitioning along the sequence dimension into partitioning along the head dimension and parallel computation of attention across different heads.
SP-Attention is a parallel version of Flash Attention~\cite{dao2023flashattention}, utilizing peer-to-peer (P2P) transmission of K and V subblock.

\begin{figure}[htbp]
\centering
\includegraphics[width=0.45\textwidth]{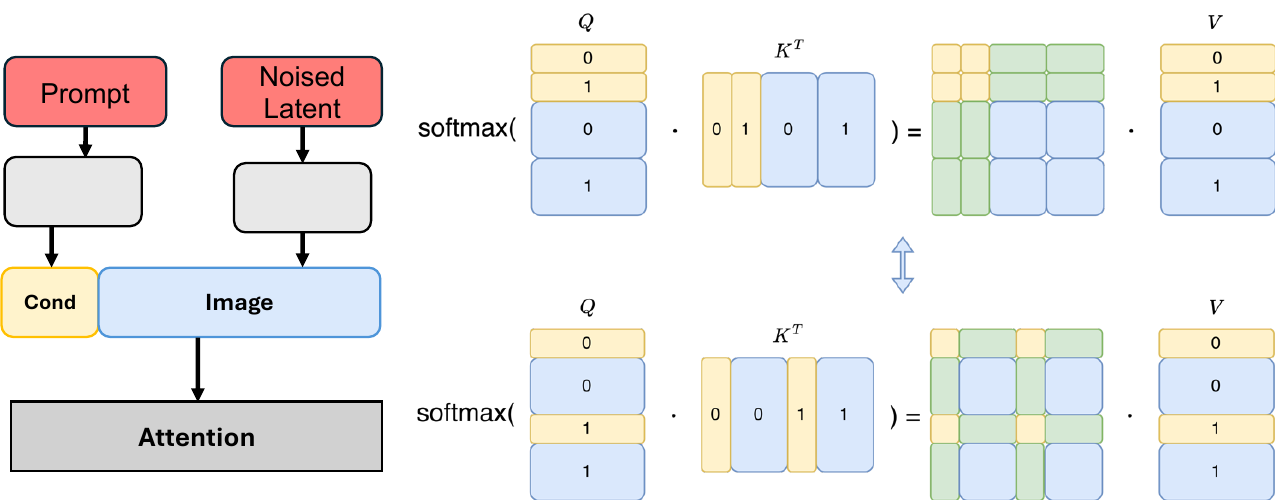}
\caption{Sequence Parallel for In-Context Conditioning applied in MM-DiTs.}
\label{fig:mmditsp}
\end{figure}

However, directly applying SP to MM-DiT Block with In-context Condition, adopted by the latest models such as SD3, Flux.1, and CogVideoX, is not feasible. As illustrated, condition information and image information are separately encoded and concatenated before Self-Attention computation. If only the sequence dimension of the image is split, the condition input tensor needs to be replicated across different devices.

We designed an SP method tailored for In-Context Condition. As shown on the right side of Figure~\ref{fig:mmditsp}, it splits both the Condition Tensor and Image Tensor along the sequence dimension. Then, it concatenates corresponding shards of condition and image input to form a local sequence. This ensures load balancing, allowing not only Attention but also the encoding logic before condition and image to be parallelized using SP. As shown, the computation yields the same results as the serial version.

\subsubsection{PipeFusion: Patch-level Pipeline Parallelism}
\label{sec:pipefusion}
Existing parallel paradigms, whether Tensor Parallel (TP) or Sequence Parallel (SP), require communication of activations for each DiTs block. 
The sequence-level pipeline parallel method proposed by TeraPipe~\cite{li2021terapipe}, used in LLMs, can transmit only limited input Activations, has been proven to be  more suitable for long sequence input~\cite{qin2024mooncake}. 
However, TeraPipe is designed for transformers using causal attention, where each token only attends to its previous tokens. In contrast, DiTs employ full attention, where each token attends to the computation of both previous and subsequent tokens.

PipeFusion~\cite{wang2024pipefusion}, a patch-level pipelined parallel approach proposed by us specifically designed for DiTs, leverages input temporal redundancy to effectively adapt TeraPipe for DiTs inference. Here, we provide a brief review of PipeFusion.

\begin{figure*}[htbp]
\centering
\includegraphics[width=0.70\textwidth]{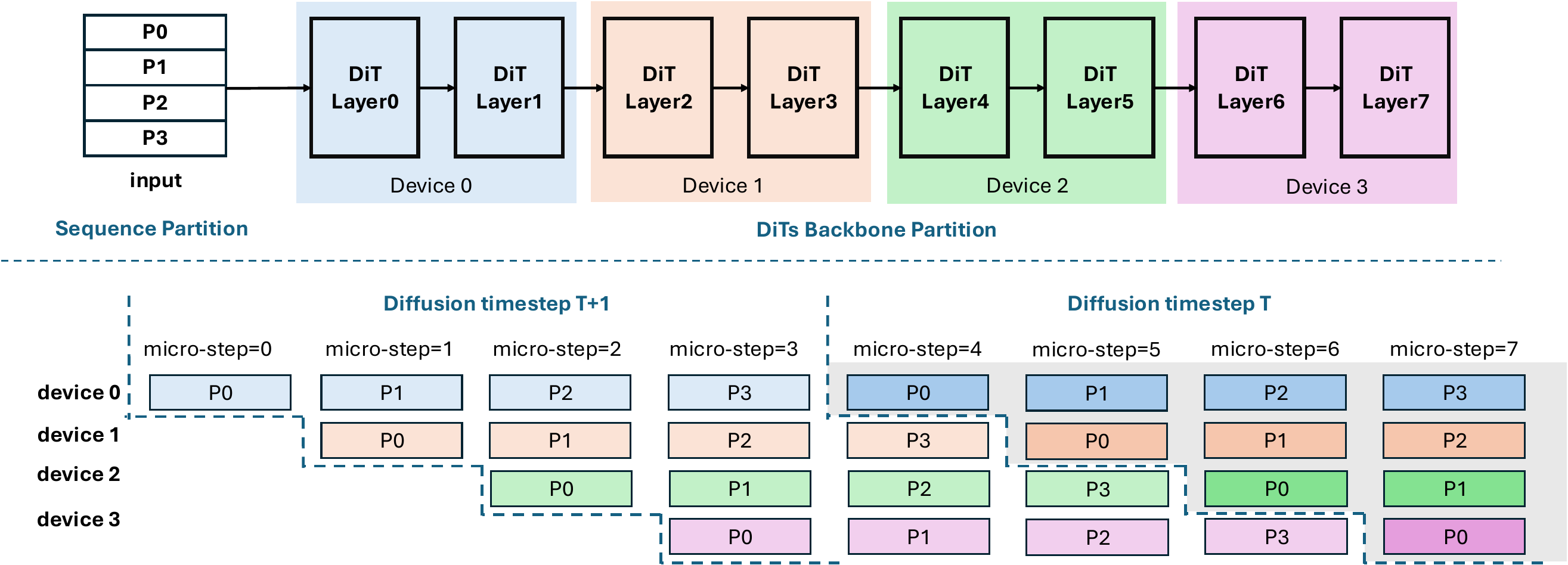}
\caption{Above: partitioning strategy for Input and DiTs backbone network. Below: Workflow of the PipeFusion as patch-level pipelined parallelism.}
\label{fig:pipefusion}
\vspace{-10pt}
\end{figure*}

PipeFusion partitions the input latent image into patches and the DiTs transformer model into layers, as shown at the top of Figure \ref{fig:pipefusion}.
It partitions the DiTs model along the data flow, assigning each partition of consecutive layers to a GPU.
The input image is also divided into 
M
M non-overlapping patches, allowing each GPU to process one patch with its assigned layers in parallel.
This pipelined approach requires synchronization between devices, which is efficient when the DiTs workload is evenly distributed across GPUs.
Achieving even partitioning is straightforward since DiTs consist of identical transformer blocks.

Suppose we are at diffusion timestep $T$, with the previous timestep being $T+1$ as the diffusion process proceeds in reverse order.
Figure~\ref{fig:pipefusion} illustrates the pipeline workflow with $N = 4$ and $M = 4$, highlighting the activation values of patches at timestep $T$.
Leveraging input temporal redundancy, a device can start its computation without waiting for full spatial activations at timestep $T$.
Instead, it uses stale activations from the previous timestep to provide context.
Additionally, in PipeFusion, devices send micro-step patch activations to subsequent devices via asynchronous P2P, enabling overlap between communication and computation.

\begin{figure}[htbp]
\centering
\includegraphics[width=0.475\textwidth]{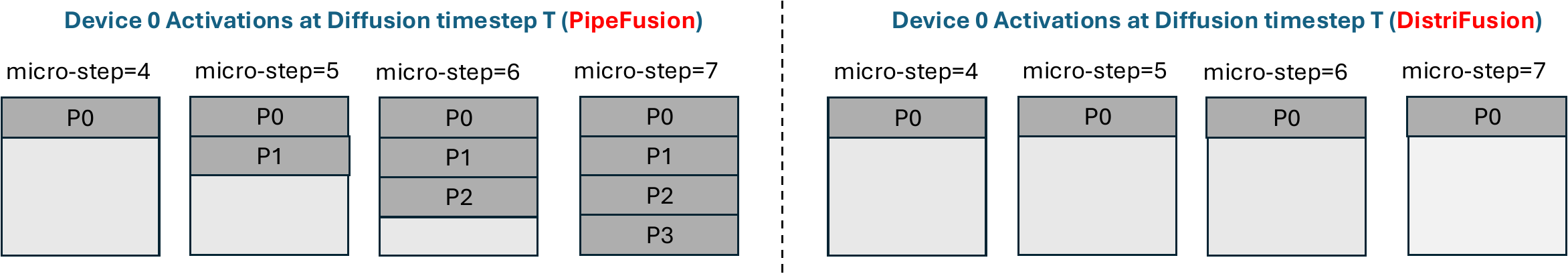}
\caption{The fresh part of activations during diffusion timestep $T$ of Figure~\ref{fig:pipefusion}. The dark gray represents fresh data and the light gray represents stable data. }
\label{fig:fresh_area}
\end{figure}

PipeFusion theoretically outperforms DistriFusion in results accuracy considering the area of fresh activation area.
As shown in Figure~\ref{fig:fresh_area}, within a single diffusion timestep, PipeFusion continuously increases the area of fresh activation as the pipeline micro-steps progress from diffusion timestep $4$ to $8$.
In contrast, throughout the entire diffusion process, DistriFusion constantly maintains one patch of fresh area out of the total $M$ patches.

Similar to DistriFusion, before executing the pipeline, we usually conduct several diffusion iterations synchronously, called warmup steps.
During the warmup phase, patches are processed sequentially, resulting in low efficiency.
Though the warmup steps cannot be executed in the pipelined manner, the workflow is relatively small compared to the entire diffusion process, and thus, the impact on performance is negligible. 

Changes in DiTs model architecture may impact PipeFusion. 
Firstly, adapting the In-Context-Conditioning to inject caption requires minor modifications to the PipeFusion algorithm. Specifically, text vectors are concatenated with Patch0, increasing its computational load and potentially causing pipeline imbalance. However, in the latest models, Flux and SD3, this impact has been found to be minimal (Figure~\ref{fig:flux_l40} and Figure~\ref{fig:sd3_16xl40}). 
For 1024px generation tasks, with image sequence lengths of 64K and 256K, and a text feature sequence of 128, the text features account for less than 2\% of the total. If the text sequence is relatively long, a load balancing image partition can be applied.
Secondly, for DiTs with skip-connection structures, a device in PipeFusion not only communicates with adjacent devices but also with a distant one, affecting P2P communication overlapping. For instance, in the 2048px task on 8$\times$A100, PipeFusion observed poor scalability due to this issue (Figure~\ref{fig:hunyuan_a100}).




\subsubsection{Comparison Between Intra-Image Parallelisms}
\label{sec:comparsion}

We analyze the communication and memory cost of existing DiTs parallel methods in Table~\ref{tab:parallel_schemes}.
The communication cost is calculated by the product of the number of elements to transfer with an \textit{algorithm bandwidth (algobw)} factor related to communication type~\footnote{https://github.com/NVIDIA/nccl-tests/blob/master/doc/PERFORMANCE.md}.
For collective algorithms of AllReduce, AllGather and AllToAll, the corresponding algobw factors are \( 2\frac{n-1}{n} \), \( \frac{n-1}{n} \), and 1. 
In the table, we approximate the term $O(\frac{n-1}{n})$ to $O(1)$ for simplicity.

\begin{table}[h]
\scriptsize
\centering
\caption{Comparison between different parallel methods for DiTs in a single diffusion timestep. Overlap denotes the overlapping between communication and computation.}
\label{tab:parallel_schemes}
\begin{tabular}{cccccc}
\toprule
\multirow{2}{*}{\textbf{Method}}& \multicolumn{2}{c}{\textbf{Communication}} & \multicolumn{2}{c}{\textbf{Memory Cost}} \\
\cmidrule(lr){2-3} \cmidrule(lr){4-5}
& \textbf{Cost} & \textbf{Overlap} & \textbf{Model} & \textbf{KV Activations} \\
\midrule
Tensor Parallelism & $4O(p \times hs)L$ & \xmark & $\frac{1}{N}P$ & $\frac{1}{N}KV$ \\

DistriFusion & $2O(p \times hs)L$ &\cmark & $P$ & $(KV)L$ \\

SP-Ring & $2O(p \times hs)L$ &\cmark & $P$ & $\frac{1}{N}KV$ \\

SP-Ulysses & $\frac{4}{N}O(p \times hs)L$ & \xmark & $P$ & $\frac{1}{N}KV$ \\

PipeFusion & 2$O(p \times hs)$ & \cmark & $\frac{1}{N}P$ & $\frac{1}{N}(KV)L$ \\
\bottomrule
\end{tabular}
\end{table}

Among all methods, PipeFusion has the lowest communication cost, as long as $N<2L$, which is easy to satisfy as the number of network layers $L$ is typically quite large, e.g., $L=38$ in Stable-Diffusion-3.
In addition, PipeFusion overlaps communication and computation.
SP-Ulysses exhibits a decreasing communication cost with increasing $N$, outperforming the remaining three methods. 
However, its communication cannot be hidden by computation.
SP-Ring and DistriFusion have similar communication costs and overlapping behaviors. 
The distinction lies in the scope of overlapping: computation in Ring Seq Parallel overlaps within the attention module, whereas that in DistriFusion overlaps throughout the entire forward pass.

To analyze the memory costs, we denote $P$ as the total number of model parameters.
In PipeFusion and tensor parallelism, the memory cost decreases as more GPUs are utilized, which is a nice property with the rapid growth of the DiT model size.
Both PipeFusion and DistriFusion maintain a KV buffer for each transformer block leading to significant activation memory overhead, especially for long sequences.
The KV buffer in PipeFusion decreases as the number of devices $N$ increases, whereas DistriFusion does not exhibit such a reduction.

Compared to PipeFusion and SP, TP and DistriFusion exhibit significant drawbacks. 
TP employs synchronous communication, with communication cost proportional to sequence length, resulting in poor scalability. Experimental results (Figure~\ref{fig:pixart_16xl40}) show that TP consistently leads to the highest latency. 
DistriFusion, on the other hand, incurs excessive memory demands for long sequences. 
It overlaps communication at the cost of increased memory usage, requiring each device to maintain communication buffers that store the complete spatial shape of K and V activations, totaling $AL$. Consequently, the memory cost of DistriFusion does not decrease with the addition of computational devices  (Figure~\ref{fig:3modelmemory}). Experimental results indicate that it is unable to infer a 0.6B Pixart model at 4096px resolution on 8$\times$L40

\subsubsection{Hybrid Parallelism}
\label{sec:hybrid_paralll}
Our experiments reveal that using SP-Ulysses, SP-Ring and PipeFusion alone sometimes is insufficient for achieving large-scale parallel inference for DiTs. 
SP-Ring, due to its higher communication overhead compared to SP-Ulysses, performs less efficiently in high-bandwidth interconnect networks. Conversely, SP-Ulysses is sensitive to hardware network topologies; for instance, The communication cost surges dramatically when All2All communication spans the QPI in PCIe-interconnected GPUs.
In this section, we propose a method to arbitrarily hybridize the three intra-image parallel methods to accommodate any network hardware topology, thereby scaling DiT inference to a large scale.
Hybrid parallel in xDiT is not as straightforward as hybrid parallel in LLMs~\cite{shoeybi2019megatron}, because PipeFusion utilizes full spatial shape KV buffers, not the KV shards belonging to itself, and the non-local K, V is difficult to update by SP correctly.

\begin{figure}[t]
\centering
\includegraphics[width=0.475\textwidth]{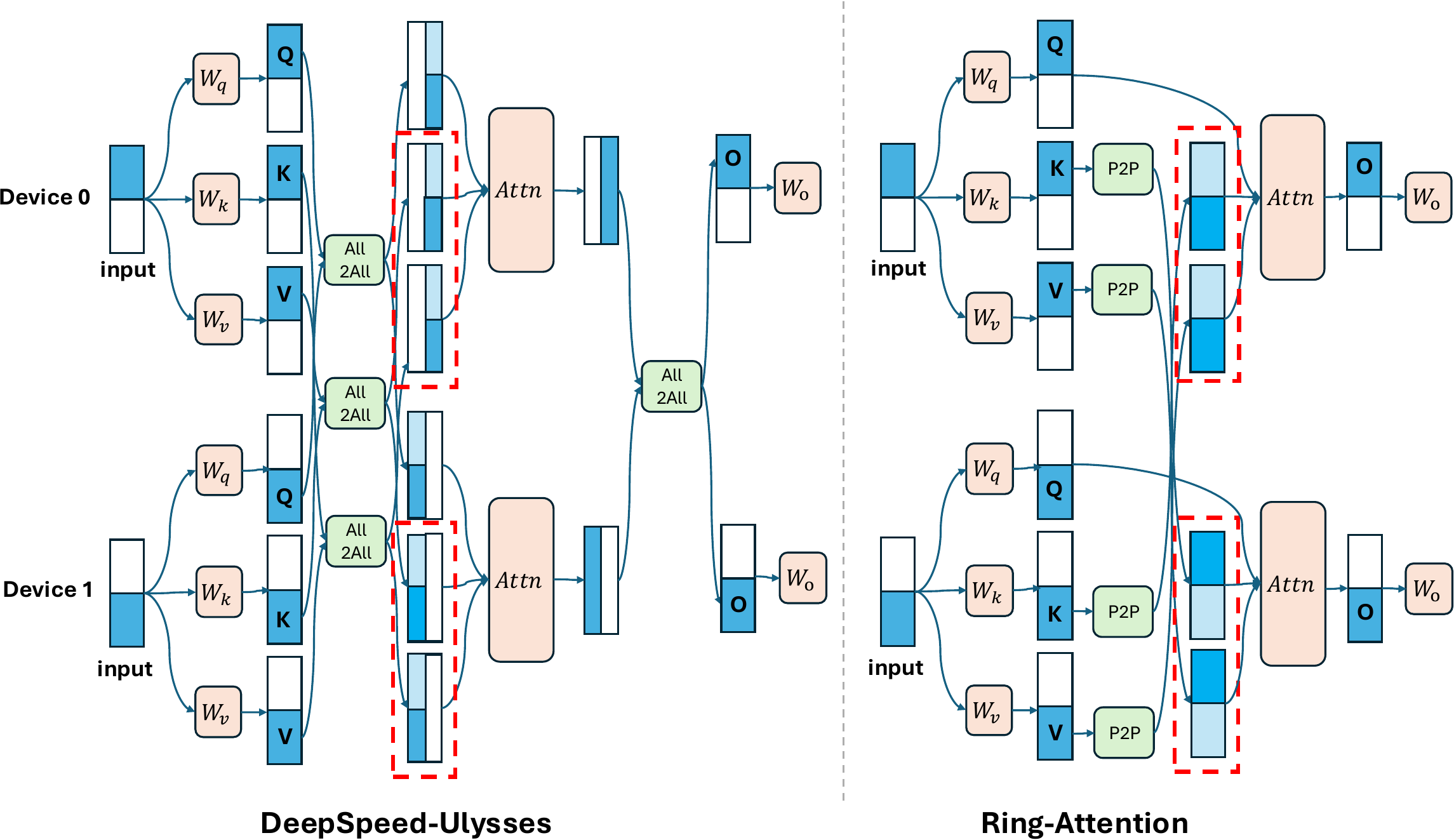}
\caption{SP-Ulysses and SP-Ring workflow. The K, V intermediate results in the red box are updated in the PipeFusion KV Buffer for SP+PipeFusion hybrid parallel.}
\label{fig:twosp}
\end{figure}

Existing research named USP~\cite{fang2024unified} has successfully hybridized SP-Ulysses and SP-Ring.
It views the process group as an 2D mesh where the columns are SP-Ring groups and rows are SP-Ulysses groups.
USP parallelizes the attention head dimension in SP-Ulysses groups and parallelizes the head dimension in SP-Ulysses groups. 
In xDiT, we need to further hybridize USP with PipeFusion.
We view the process group for intra-image parallelism as a 2D mesh of pipefusion\_degree $\times$ sp\_degree. 
PipeFusion is executed on the high dimension, while the USP is executed on the lower dimension. 
Both SP and PipeFusion split the input palong the sequence dimension. The entire hidden state is first split into M patches along the sequence dimension, and each patch is further split into sp\_degree patches. As illustrated in the figure, 8 devices are divided into pipefusion\_degree=4, sp\_degree=2, and $M$=4.

\begin{figure}[t]
\centering
\includegraphics[width=0.45\textwidth]{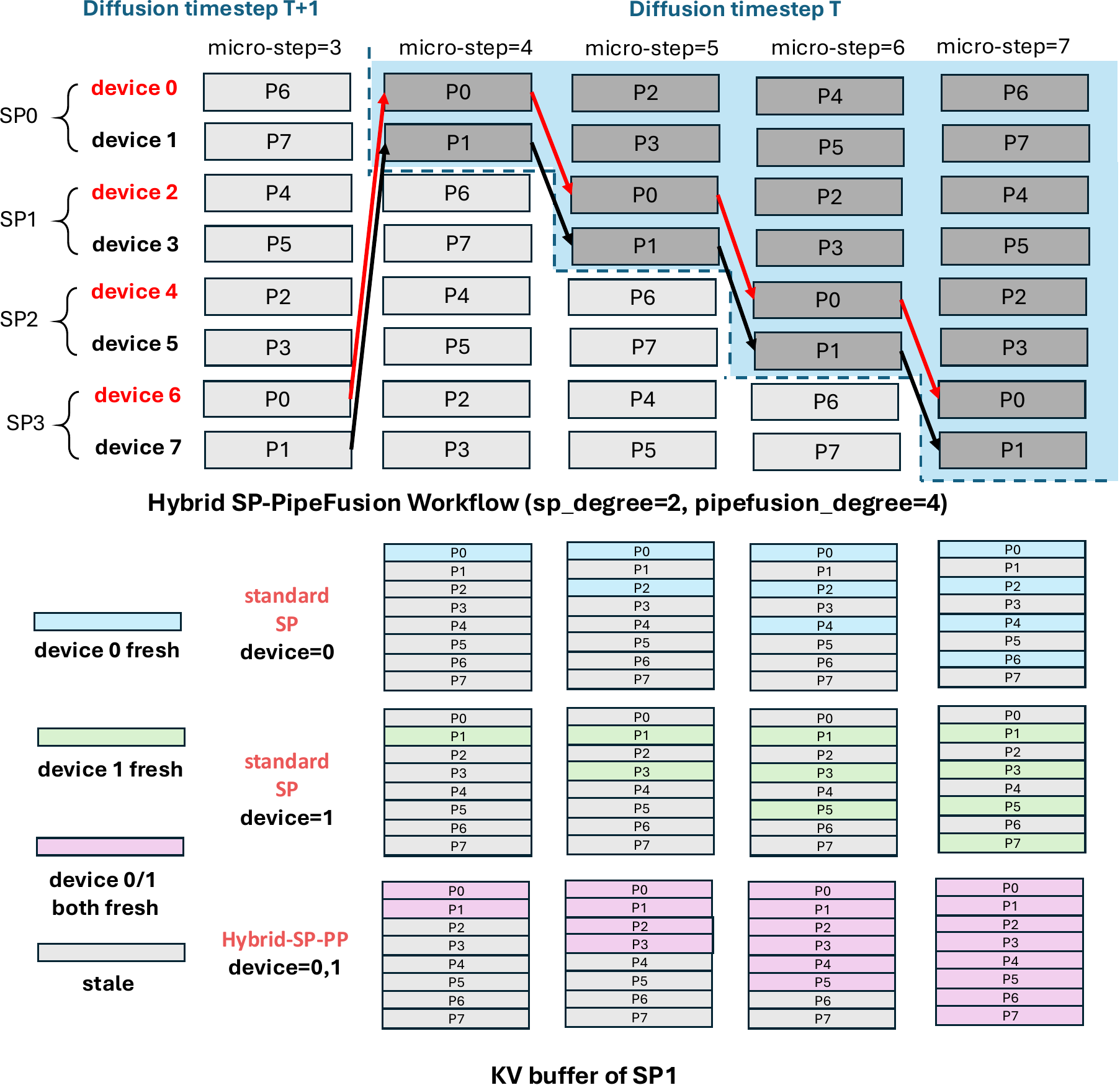}
\caption{Hybrid PipeFusion and Sequence Parallel Workflow.}
\label{fig:hybrid_parallel}
\end{figure}

The challenge of hybrid parallel combining PipeFusion and SP lies in correctly updating the K, V of attention. If PipeFusion and SP are naively computed within their respective process groups, PipeFusion will use incorrect stale K, V values. As marked "standard SP" in the lower part of Figure~\ref{fig:hybrid_parallel}, device 0 only updates K, V belonging to the even-numbered patches, while device 1 only updates K, V belonging to the odd-numbered patches. 
In this case, this results in half of the KV on each device not being updated with the desirable K, V values from the previous diffusion step.

The key to design correct SP+PipeFusion hybrid parallel is that the KV involved in Attention computation on different devices within the SP group should be consistent with each other. 
We designed a highly elegant method to achieve this without introducing any overhead, requiring only minor modifications to the SP algorithm. 
As shown in Figure~\ref{fig:twosp}, after communication of K and V in SP-Ulysses and SP-Ring, the intermediate results (in the dashed red box) are stored in each device's KV Buffer. 
However, in the standard SP implementations for both SP-Ring and SP-Ulysses, these intermediate results are discarded after the Attention computation. 
As shown in the "Hybrid-SP-PP" part of Figure~\ref{fig:hybrid_parallel}, the devices 0 and 1 in the same SP process group are updated with consistent K, V values.
For SP-Ulysses, we obtain the KV of the sequence within the SP group participating in the computation of the head, and we do not need information from non-participating heads. 
For SP-Ring, we obtain the KV of the sequence within the SP group for all heads.

\subsection{Inter-Image Parallelism}
\label{sec:inter-image-parallel}
The Classifier-Free Guidance (CFG)~\cite{ho2022classifier} has become an important optimization for diffusion models by providing broader conditional control, reducing training cost, enhancing the quality and details of generated content, and improving the practicality and adaptability of the model.
For an input prompt, using CFG requires generating both unconditional guide and text guide simultaneously, which is equivalent to inputting input latents batch size=2 of DiT blocks. 

CFG parallelism separates the two latents for computation, and after each diffusion step forward is completed and before the scheduler executes, it performs an Allgather operation on the latent space results. Its communication overhead is much smaller than PipeFusion and sequence parallelism. Therefore, when using CFG, CFG parallelism must be used.

\subsection{Parallel VAE}
\label{sec:vae}
After the DiT backbone denoised an image in latent space, an autodecoder module employs a VAE~\cite{kingma2013auto} to decode the image (size $\frac{h}{8}\times \frac{w}{8} \times c$), where $c$ is the channel size and usual be set as 4 or 16, from the latent space into images in the pixel space ($h\times w \times 3$). 
The VAE applies multiple convolutional neural networks for upsampling. 
Unfortunately, as the height and width dimensions of the feature maps increase, especially after passing through convolutional layers with numerous channels, two significant issues emerge. 
Firstly, there is a substantial increase in activation memory. 
For example, in 4096px image generation, the peak activation tensor in the SD-VAE~\footnote{https://huggingface.co/stabilityai/sd-vae-ft-mse} reaches 60.41 GB.
Secondly, excessive temporary memory for convolutional operators of large inputs also leads to memory spikes.
This memory bottleneck hinders users and the research community from scaling up models to produce high-quality images.

To address the activations OOM issue, xDiT applies patch parallelism, the sequence parallelism, named as patch parallelism here, for 2D inputs, for the VAE modules.
We divide the feature maps in the latent space into multiple patches and perform parallel VAE decoding across different devices.
This requires the exchange of the boundary data for convolutional operators by allgther communications.
Patch parallelism reduces the peak memory for intermediate activations to $\frac{1}{N}$.
While VAE parameters are replicated across devices, their total size is relatively small (320MB in this case).
To tackle the temporal memory spike issue, previous research~\cite{hanlab2023patchconv, yang2024cogvideox} proposed decomposing the execution of a single convolutional operator into multiple stages, processing portions of the input data sequentially to minimize temporary memory consumption. By combining these two methods, xDiT's VAE can generate an image resolution of 7168px, more than 12.25 times larger than using the naive VAE approach on 8$\times$L40 (48GB).
The detailed results are presented in Sec.~\ref{sec:parallvaeresults}.

\section{Experiments}

\subsection{Setups}

In our experimental setup, we evaluate xDiT on two GPU clusters.
The first cluster contains two nodes, each equipped with 8$\times$L40-48GB (PCIe Gen4$\times$16), connected via a 100Gbps Ethernet network.
The second cluster consists of a single node having 8$\times$A100-80GB (NVLink).

\begin{table}[t]
\Small
\centering
\begin{tabular}{|c|c|c|c|}
\hline
\textbf{Model} & \textbf{Transformers} & \textbf{Text-Encoder} & \textbf{VAE} \\
\hline
Pixart~\cite{chen2023pixart, chen2024pixart} & 2.3GB (0.6B) & 18GB & 320MB \\
\hline
SD3-medium~\cite{esser2024scaling} & 7.8GB (2B) & 19GB & 320MB \\
\hline
Flux.1-dev~\cite{blackforestlabs2023announcement} & 23GB (12B) & 9.1GB & 320MB \\
\hline
HunyuanDiT~\cite{li2024hunyuan} & 5.6GB (1.5B) & 7.7GB & 320MB \\
\hline
CogVideoX-5B~\cite{yang2024cogvideox} & 11GB (5B) & 8.9GB & 412MB \\
\hline
\end{tabular}
\caption{Disk Usage for Different Components of four DiTs. The Transformers entries include the number of parameters.}
\label{tab:dit_models}
\end{table}

Our analysis involves five DiTs as shown in Table~\ref{tab:dit_models}.
Official weights from HuggingFace were utilized across all models. 
Among these, the CogVideoX-5B model focuses on video generation, while the remaining models are oriented towards image generation.
Notably, Pixart and HunyuanDiT leverage cross-attention for conditioning injection within DiT blocks, whereas the other models combine text and image modality encoding information through in-context conditioning. 
In real applications, the VAE can operate independently on separate computing resources, so we evaluate the performance of the backbone (text encoder+transformers) and the VAE components separately.

For PipeFusion and DistriFusion, we both apply 1 step warmup iteration. PipeFusion selects the best latency performance by searching the patch number $M$ from 2, 4, 8, 16, 32. For each experiment, we show the average latency of 5 runs.







\subsection{Backbone Performance}

\subsubsection{Results on two 8$\times$L40 nodes}
In this section, we present the scalability of various models on two 8$\times$L40 nodes connected by 100Gbps Ethernet.
Besides studying every single method introduced by our system, we also explore the optimal hybrid parallel methods. 
Furthermore, we compare existing techniques including tensor parallelism and DistriFusion on the Pixar model to gain systematic insight.

\textbf{Pixart:}
Figure~\ref{fig:pixart_a100} illustrates the scalability of various parallel strategies on Pixart for generating image resolutions of 1024px, 2048px, and 4096px.
The Pixart model, capable of generating high-resolution images (up to 4096px), is the most similar to the original first version of DiTs~\cite{peebles2023scalable}.

Applying a single parallel method leads to notably poor performance on 16$\times$L40
For TP, SP-Ulysses, SP-Ring, and DistriFusion, the latency on 16 GPUs is several times that on 8 GPUs.
This disparity stems from the substantial communications volumes in AllReduce, All2All, AllGather and P2P communication, which exhibit high latency across Ethernet-connected nodes. By contrast, using CFG parallelism for inter-node communication proves crucial since it only performs an AllGather of the latent image once after a diffusion step, resulting in decreased communication compared to intra-image parallelism.

\begin{figure}[b]
\centering
\includegraphics[width=0.475\textwidth]{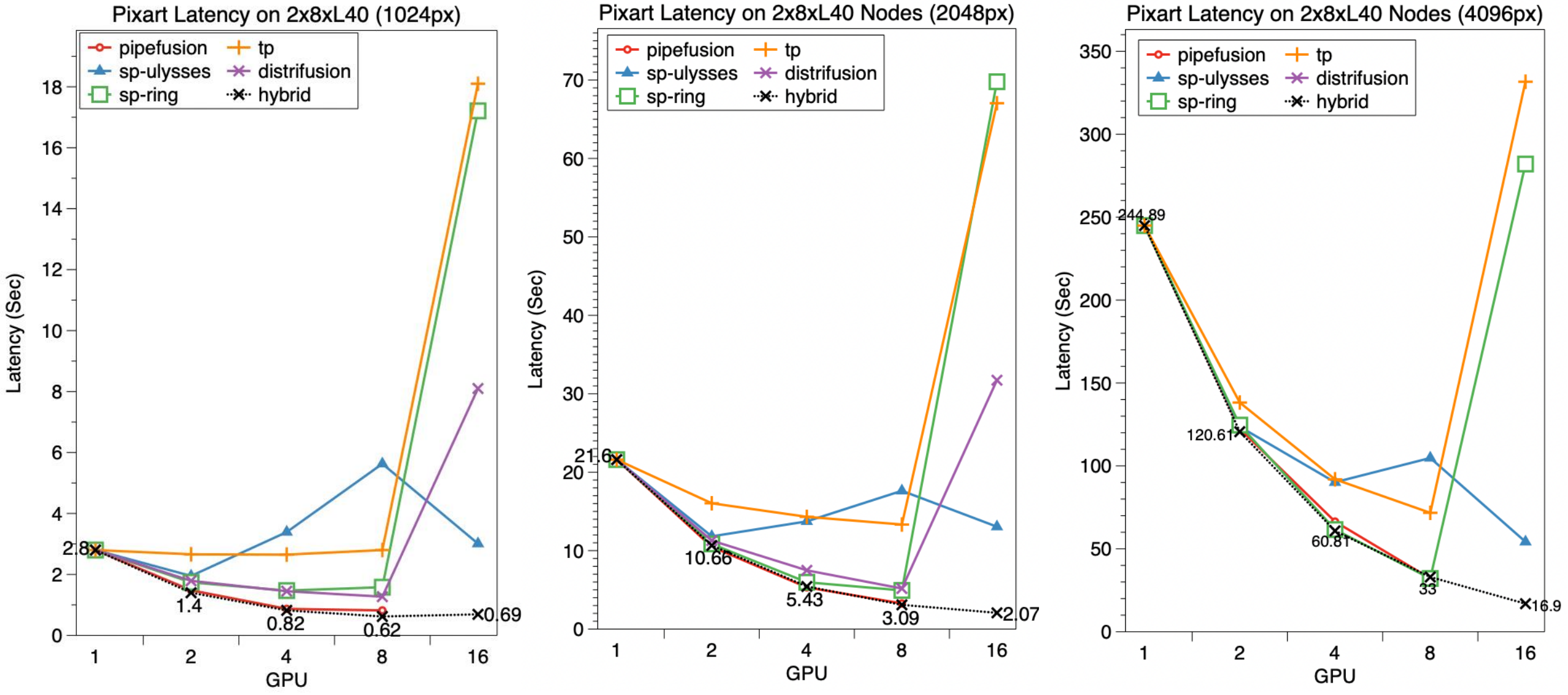}
\caption{Scalability of various parallel approaches on 2$\times$8$\times$L40 GPUs for Pixart image generation tasks employing the 20-Step DPM scheduler.}
\label{fig:pxiart-l40}
\end{figure}

\begin{figure}[t]
\centering
\includegraphics[width=0.38\textwidth]{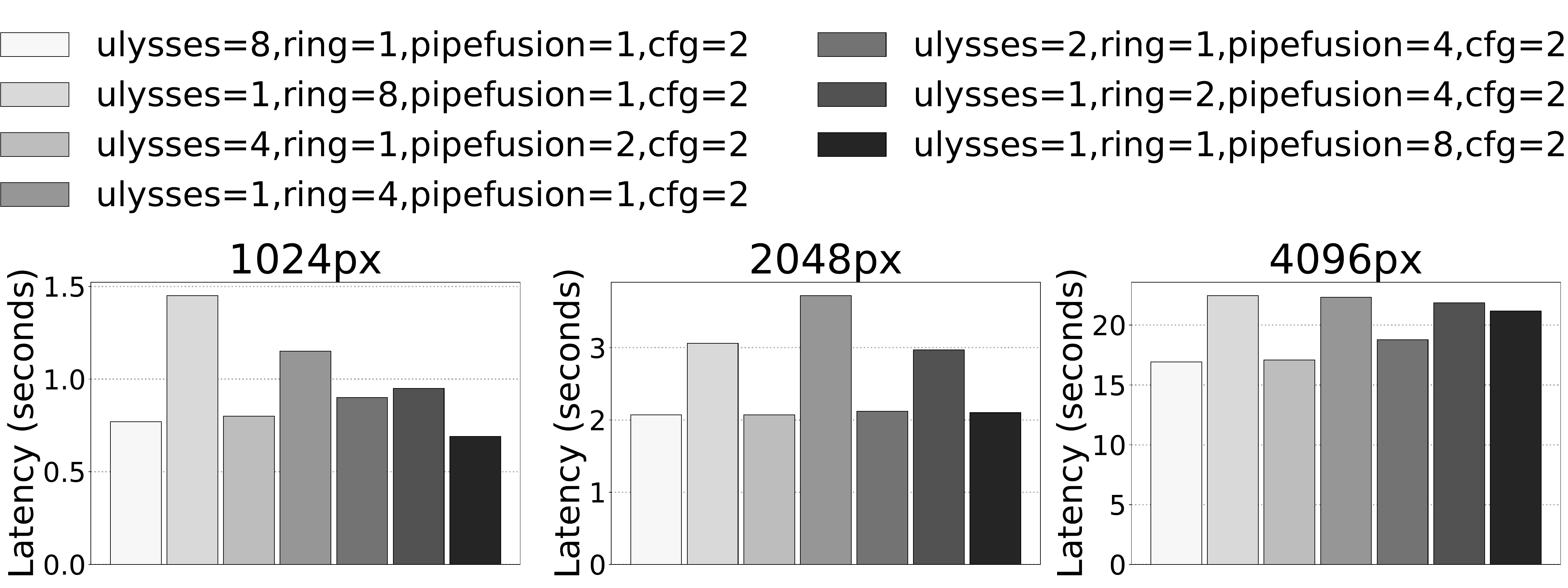}
\caption{Latency of different hybrid parallel configurations on 2$\times$8$\times$L40 GPUs for Pixart employing the 20-Step DPM scheduler.}
\label{fig:pixart_16xl40}
\end{figure}

The 0.6B Pixart model has the lowest computational load among all target models, making it the most challenging to scale.
Despite this, \textbf{\textit{xDiT achieved a perfect speedup of 13.29$\times$ for the 4096px task on 16 GPUs, lowering the latency from 245 sec to 17 sec}}!
Figure \ref{fig:pixart_16xl40} further shows the latency across hybrid parallel configurations for Pixart on 16$\times$L40. The best hybrid parallel strategy varies on different image sizes. 
For the 1024px task, the lowest diffusion latency was achieved with pipefusion=8 and cfg=2. For the 2048px task, both ulysses=4, pipefusion=2, cfg=2 and ulysses=4, cfg=2 emerged as best configuration. For the 4096px task, the best hybrid parallel configuration was ulysses=8, cfg=2. 


For the 4096px task, DistriFusion encounters OOM issues due to the linearly increasing memory cost of the KV Buffer with the sequence length.
Given that DistriFusion is only suitable for GPUs with ample memory, we preclude DistriFusion from serving as the fundamental parallel method in xDiT. 
PipeFusion encountered NCCL timeout issues when scaled to 16 GPUs, resulting in missing data.

When scaling within a single node of 8$\times$L40, PipeFusion demonstrated significantly lower latency compared to other methods on 8 GPUs. 
DistriFusion is the second-best parallel method.
In contrast, SP-Ulysses, SP-Ring, and tensor parallelism showed higher latency compared to 4$\times$L40 due to increased communication overhead across the CPU's QPI bus on PCIe Gen4. Nonetheless, for image generation tasks at various resolutions and on different numbers of GPUs, the optimal parallel strategy consistently involved a hybrid of multiple parallel methods.

\begin{figure}[t]
\centering
\includegraphics[width=0.475\textwidth]{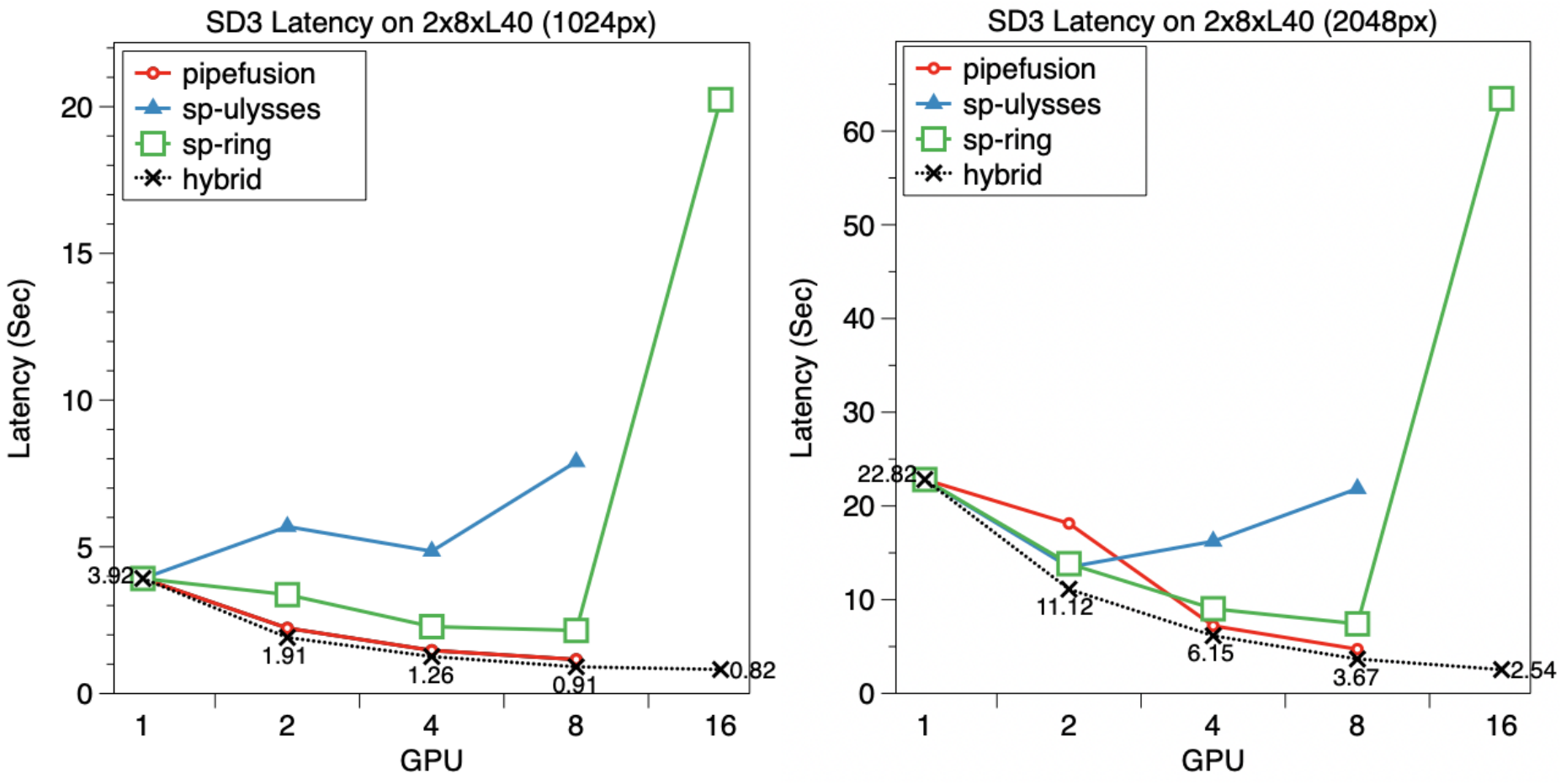}
\caption{Scalability of various parallel approaches on 2$\times$8$\times$L40 GPUs for SD3 image generation tasks employing the 20-Step FlowMatchEulerDiscrete scheduler.}
\label{fig:sd3-l40}
\end{figure}

\begin{figure}[t]
\centering
\includegraphics[width=0.38\textwidth]{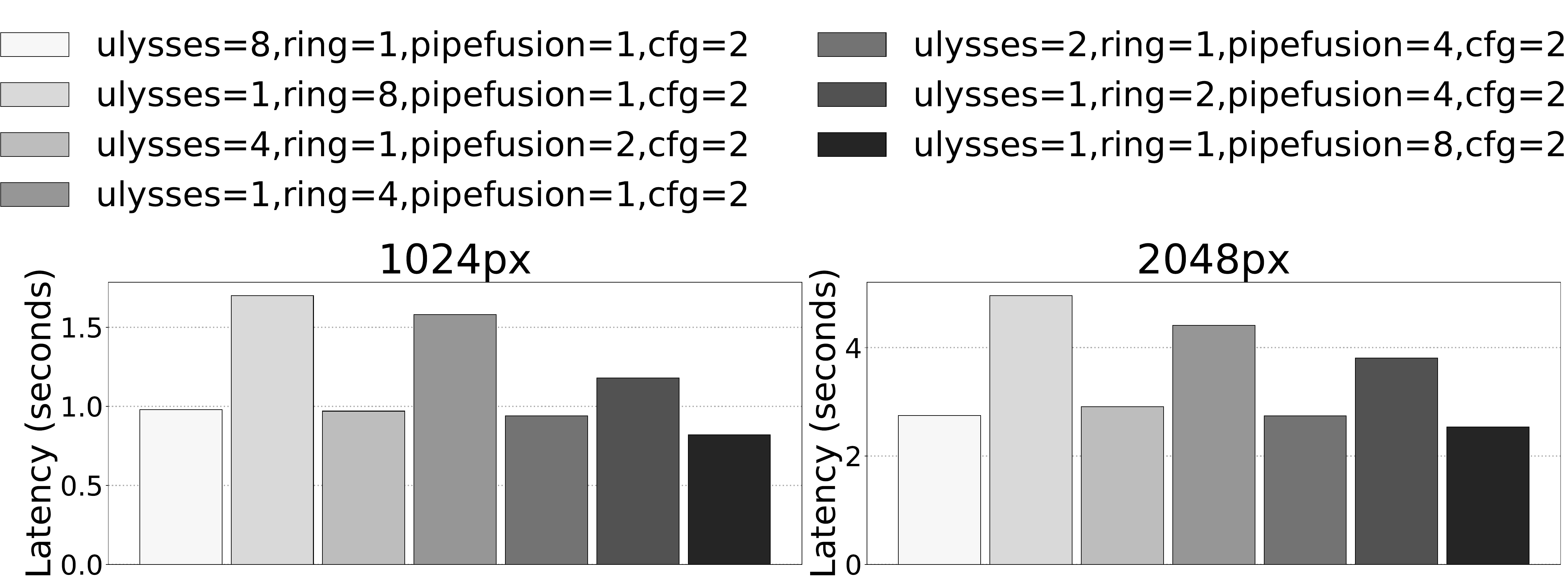}
\caption{Latency of various parallel approaches on 2$\times$8$\times$L40 GPUs for SD3 image generation tasks employing the 20-Step FlowMatchEulerDiscrete scheduler.}
\label{fig:sd3_16xl40}
\end{figure}

\textbf{SD3:}
In evaluations for SD3, we exclude tensor parallelism and Distrifusion because of their significant time and memory inefficiencies. 
We study the scalability of SD3 on two 8$\times$L40 nodes. 
Due to the constraint on the number of attention heads, 16 does not divide evenly into 24, thus preventing the application of SP-Ulysses with a degree parallel of 16 to SD3.
The remaining configurations exhibit a similar trend to Pixart, as demonstrated in Figure~\ref{fig:sd3-l40}. 
Specifically, as we increase the number of GPUs from 8 to 16, SP-Ring meets a significant performance decline, while the hybrid parallel still demonstrates further speedup. Figure \ref{fig:sd3_16xl40} details the performance of SD3 on 16$\times$L40. For both the 1024px and the 2048px tasks, the strategy with pipefusion=8 and cfg=2 displayed the shortest latency, while the configuration with pipefusion=4, ulysses=2, and cfg=2 ranked as the second fastest.

\begin{figure}[b]
\centering
\includegraphics[width=0.48\textwidth]{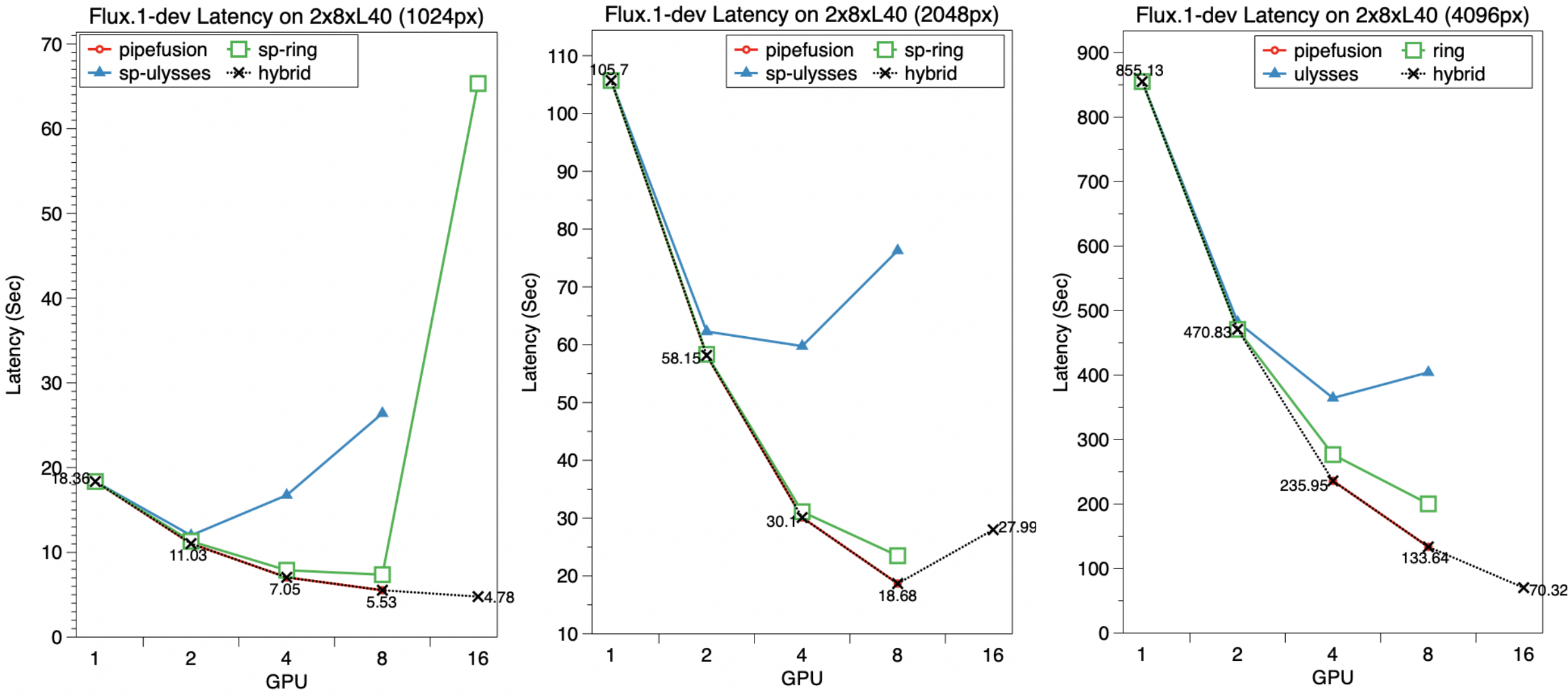}
\caption{Scalability of various parallel approaches on 2$\times$8$\times$L40 GPUs for Flux.1-dev image generation tasks employing the 28-Step FlowMatchEulerDiscrete scheduler.}
\label{fig:flux_l40}
\end{figure}

\textbf{Flux.1-dev:}
Figure~\ref{fig:flux_l40} shows the scalability of Flux.1 on two 8$\times$L40 Nodes. 
For the same reason as SD3, SP-Ulysses with a parallel degree of 16 cannot be applied to Flux.1.
As Flux.1 does not use the CFG technique, CFG parallel is not applicable. 
However, we can still achieve enhanced scalability by using PipeFusion as a method for parallel between nodes.
For the 1024px task, hybrid parallel on 16$\times$L40 is 1.16x lower than on 8$\times$L40, where the best configuration is ulysses=4 and pipefusion=4.
For the 4096px task, hybrid parallel on 16$\times$L40 is 1.9x lower than on 8$\times$L40, where the best configuration is ulysses=2, ring=2, and pipefusion=4.
However, the performance improvement is not achieved with 16 GPUs on 2048px tasks.



\textbf{CogVideoX:}
CogVideoX-5B generates 49 frames of 720x480 video using the 50-Step DDIM scheduler. PipeFusion has not yet been applied due to the distinct Temporal Redundancy characteristics~\cite{zhao2024real, meta2024moviegen} of video models compared to image models and the lack of comprehensive research in this area. xDiT employs a hybrid parallel of SP and CFG parallel.

\begin{figure}[t]
\centering
\includegraphics[width=0.35\textwidth]{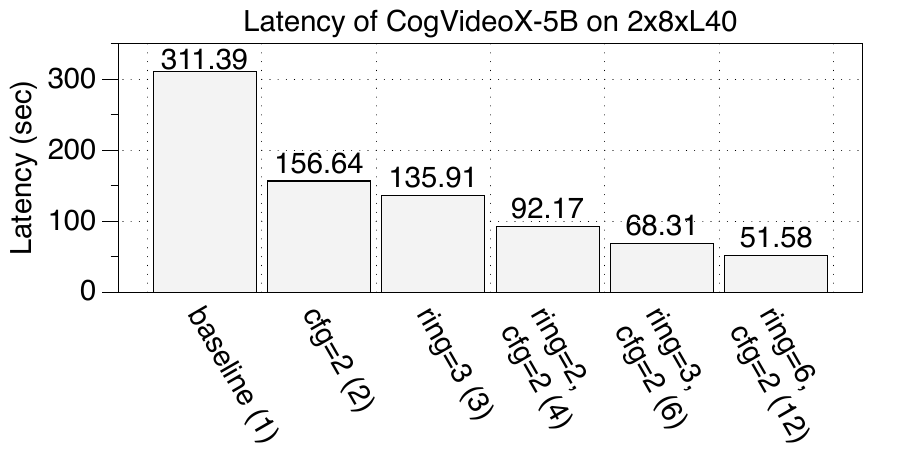}
\caption{Latency of best hybrid parallel for various parallel degrees on 2$\times$8$\times$L40 GPUs for CogVideoX-5B employing the 50-Step DDIM scheduler.}
\label{fig:cogvideo_16xl40}
\end{figure}

The best hybrid parallel configurations for different degrees of parallel are illustrated in Figure~\ref{fig:cogvideo_16xl40}.
Due to the height=480 limitation, SP-Ring cannot scale to 8$\times$GPUs. Due to the number of attention heads=30 constraint, SP-Ulysses cannot scale to 4$\times$GPUs. On 2$\times$L40 nodes, it can at most scale up to 2$\times$6 GPUs. 
xDiT achieved a 6.0$\times$ speedup across 12 L40s connected via Ethernet and a 4.55$\times$ speedup across 6 L40s interconnected via PCIe. 
\textit{\textbf{xDiT successfully reduced the video generation latency from over five minutes to just 52 seconds.}}

\subsubsection{Results on one 8$\times$A100 Node}

This section presents latency results for DiT backnone on an 8$\times$A100 (NVLinK) node.
The NVLink provides 600 GB/s bandwidth for any two GPUs.
This is 48 times the 100 Gbps bi-section Ethernet bandwidth and 18.75 times the PCIe Gen4 bus bandwidth.

\begin{figure}[b]
\centering
\includegraphics[width=0.475\textwidth]{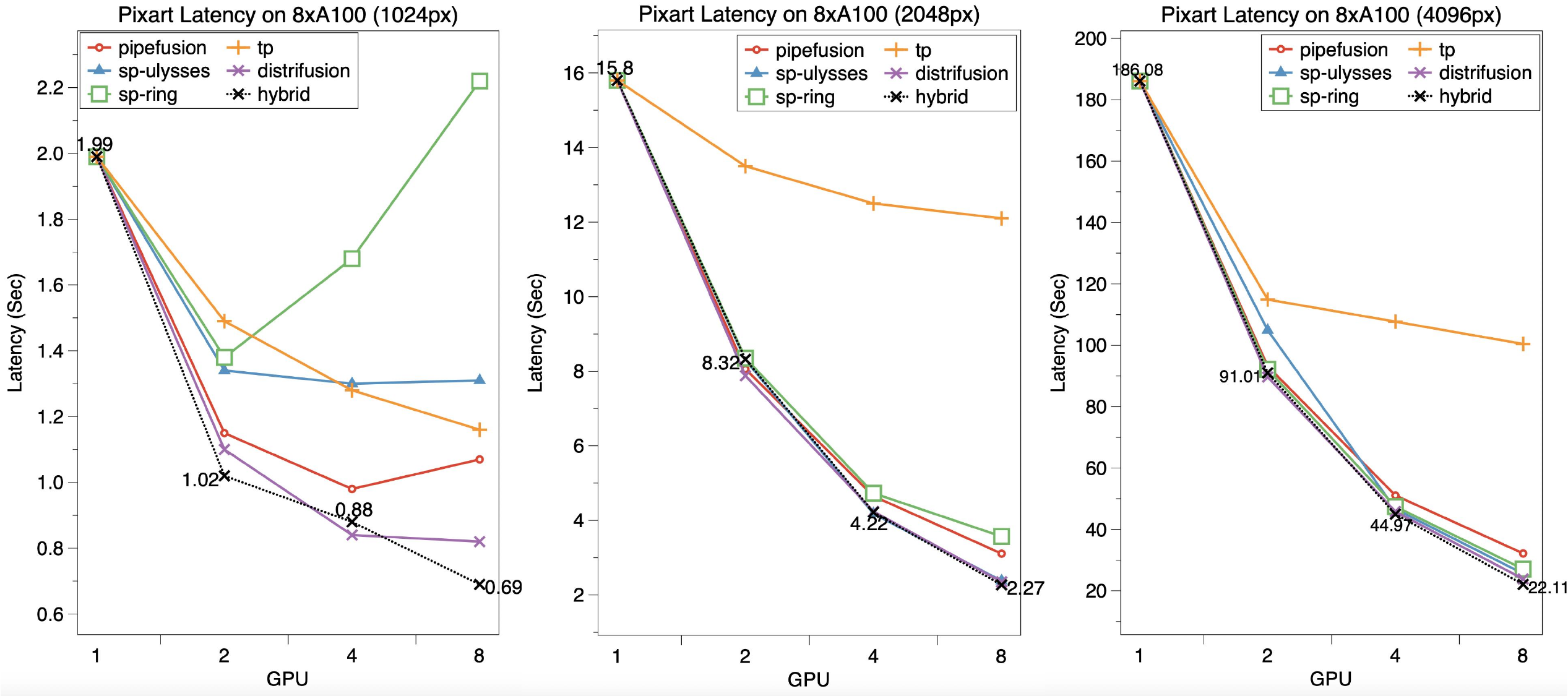}
\caption{Scalability of various parallel approaches on 8$\times$A100 GPUs for Pixart image generation tasks employing the 20-Step DPM scheduler.}
\label{fig:pixart_a100}
\end{figure}

\textbf{Pixart:}
Figure~\ref{fig:pixart_a100} presents the scalability of Pixart on 8$\times$A100 GPUs.
For the 4096px task, xDiT achieved linear speedup, \textit{\textbf{with speedup of 8.4$\times$, 4.1$\times$, and 2.0$\times$, compared to the 1 GPU baseline}}. 
On 8 GPUs, the latency decreased from 186 seconds to 22 seconds, with communication cost perfectly eliminated.

For the 1024px task, the optimal xDiT hybrid parallel configurations are as follows: cfg=2 for 2 GPUs, cfg=2 and pipefusion=2 for 4 GPUs, cfg=2 and pipefusion=4 for 8 GPUs.
DistriFusion and PipeFusion as individual parallel methods also exhibit favorable latencies.
SP-Ulysses and tensor parallelism show limited scalability.
SP-Ring performance degrades with 4 and 8 GPUs due to high communication cost.

For the 2048px image generation task, the optimal xDiT hybrid parallel configurations are as follows: cfg=2 for 2 GPUs, cfg=2 or ulysses=2 for 4 GPUs, cfg=2 and ulysses=4 for 8 GPUs.
DistriFusion, SP-Ulysses, and hybrid parallelism exhibit similar scalability, outperforming PipeFusion, SP-Ring, and tensor parallelism. Specifically, hybrid parallelism exhibits 1.03x and 1.0x lower latency compared to DistriFusion on 4 and 8 GPUs, respectively.

For the 4096px image generation task, the optimal hybrid parallel configurations remain the same as for the 2048px task. In this scenario, DistriFusion, SP-Ulysses, SP-Ring, and hybrid parallelism demonstrate comparable scalability. However, hybrid parallelism achieves slightly lower latency than DistriFusion, with a 1.02x and 1.08$\times$ speedup on 4 and 8 GPUs, respectively. PipeFusion underperforms relative to SP-Ring, while tensor parallelism exhibits the poorest performance across all methods.

\begin{figure}[t]
\centering
\includegraphics[width=0.475\textwidth]{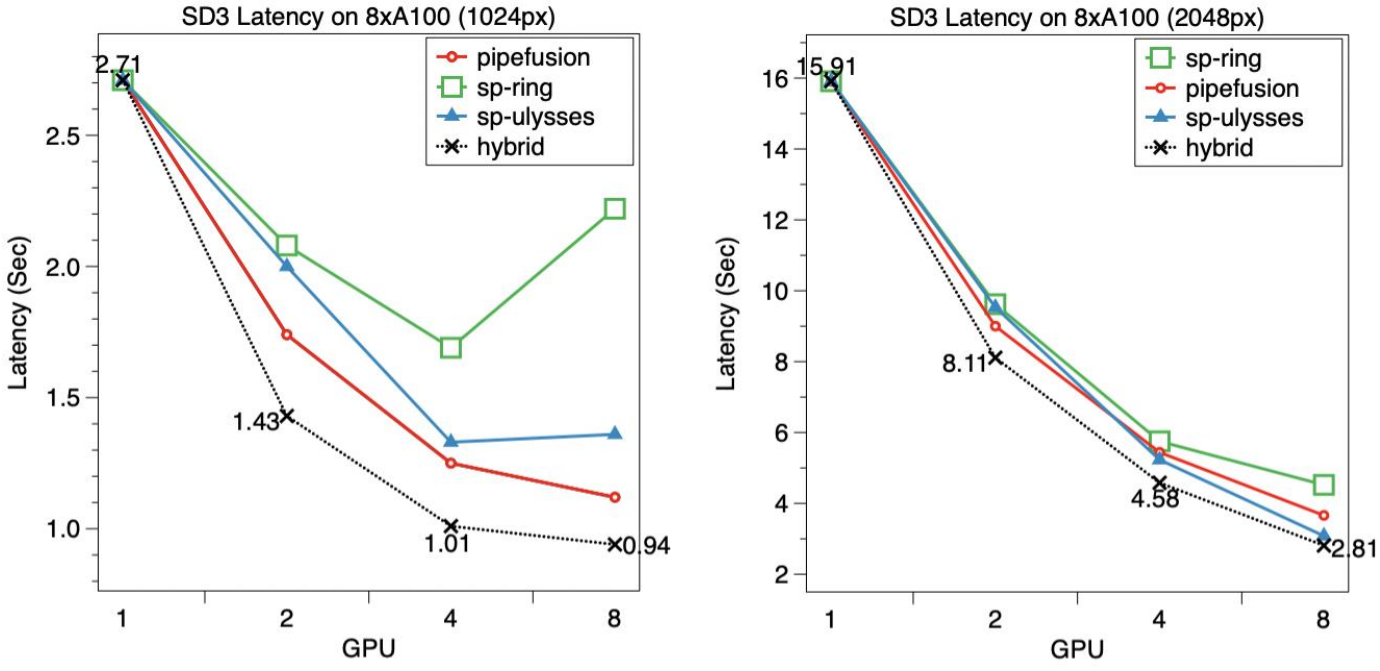}
\caption{Scalability of various parallel approaches on 8$\times$A100 GPUs for SD3-medium image generation tasks employing the 20-Step FlowMatchEulerDiscrete scheduler.}
\label{fig:sd3_a100}
\end{figure}

\textbf{SD3:}
Figure~\ref{fig:sd3_a100} illustrates the scalability of SD3 on 8$\times$A100 GPUs.
For the 1024px image generation task, PipeFusion consistently provides the lowest latency when applied as a single parallel method.
The optimal approach for hybrid parallel in this case is cfg=2, with the remaining parallel degree allocated to PipeFusion.

For the 2048px image generation task, PipeFusion achieves the lowest latency on 2 GPUs, whereas SP-Ulysses provides the lowest latency on 4 and 8 GPUs.
The optimal approach for hybrid parallelism is cfg=2, with the remaining parallelism allocated to SP-Ulysses.

\begin{figure}[htbp!]
\centering
\includegraphics[width=0.475\textwidth]{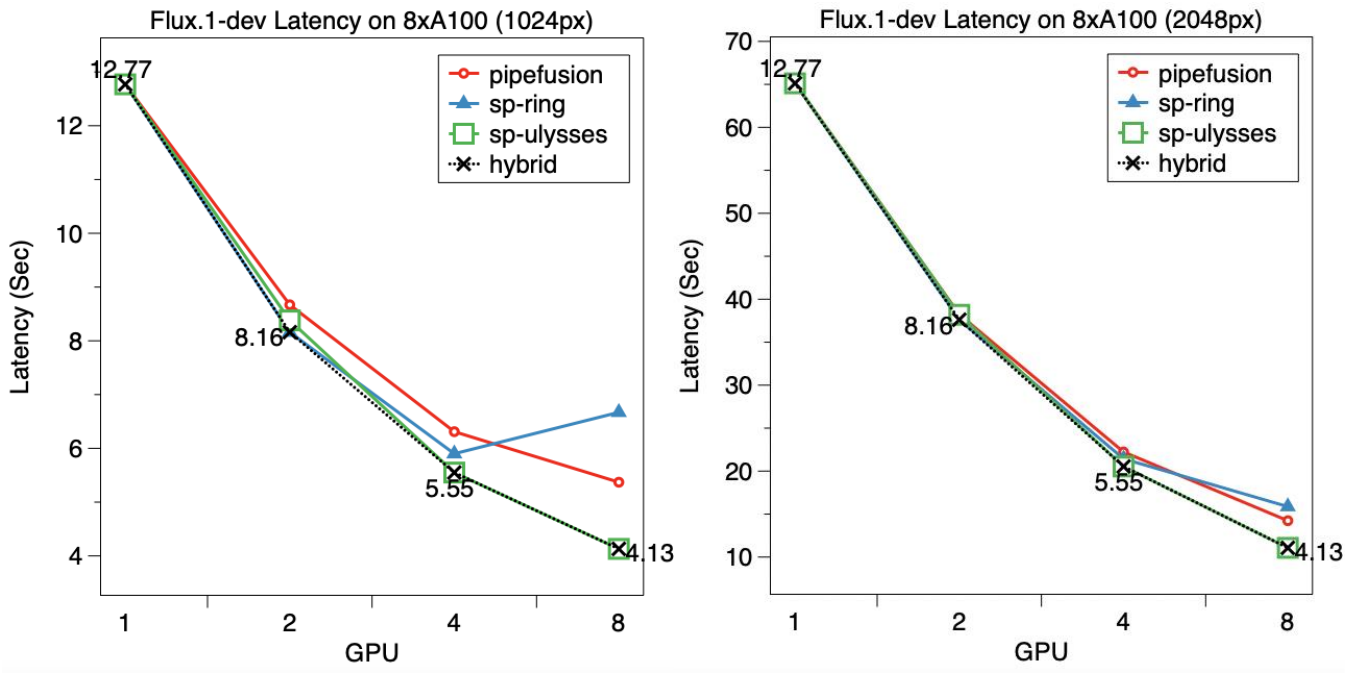}
\caption{Scalability of various parallel approaches on 8$\times$A100 GPUs for Flux.1-dev image generation tasks employing the 28-Step FlowMatchEulerDiscrete scheduler.}
\label{fig:flux_a100}
\end{figure}

\textbf{Flux.1-dev:}
Figure~\ref{fig:flux_a100} demonstrates the scalability of Flux.1 on 8$\times$A100 GPUs.
For both the 1024px and the 2048px image generation tasks, SP-Ulysses exhibits the lowest latency among the single parallel methods. The optimal hybrid strategy also incorporates SP-Ulysses in this case.

\begin{figure}[t]
\centering
\includegraphics[width=0.475\textwidth]{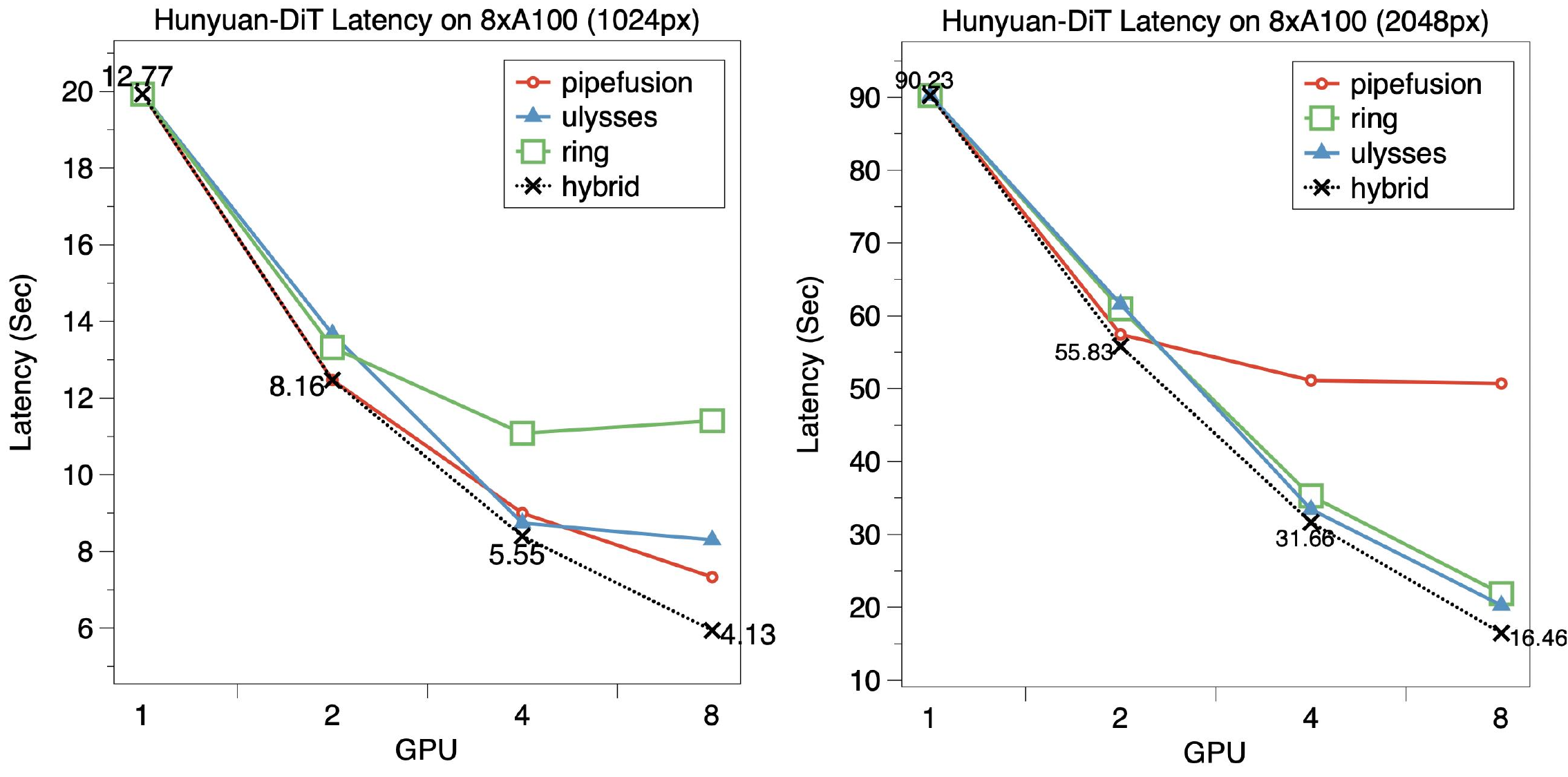}
\caption{Scalability of various parallel approaches on 8$\times$A100 GPUs for HunyuanDiT image generation tasks employing the 50-Step DPM scheduler.}
\label{fig:hunyuan_a100}
\end{figure}

\textbf{HunyuanDiT:}
Figure~\ref{fig:hunyuan_a100} illustrates the scalability of HunyuanDiT on 8$\times$A100 GPUs. HunyuanDiT employs DiT Blocks interconnected via Skip Connections, with each DiT Block connected to both an adjacent and a non-adjacent block.

\begin{figure*}[t]
\centering
\includegraphics[width=0.82\textwidth]{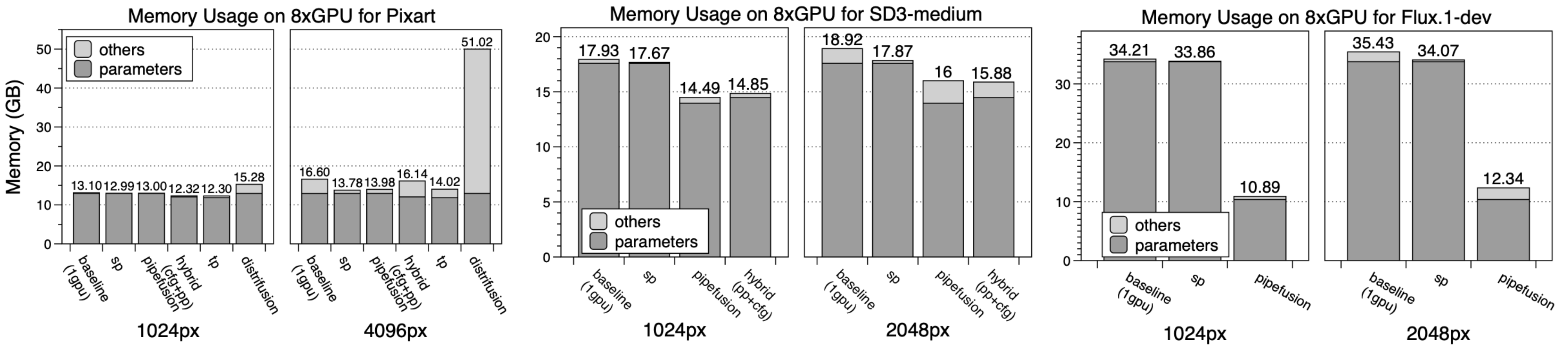}
\caption{Max GPU Memory Usage of Various Parallel Approaches.}
\label{fig:3modelmemory}
\end{figure*}

For the 1024px image generation task, the optimal hybrid parallel configurations are as follows: pipefusion=2 for 2 GPUs; cfg=2, pipefusion=2 for 4 GPUs; cfg=2, pipefusion=4 for 8 GPUs.
Hybrid parallelism achieves a 1.04x and 1.23x speedup over single parallel methods on 4 and 8 GPUs, respectively. PipeFusion exhibits lower latency than SP-Ulysses on 8 GPUs but similar latency on 4 GPUs. 
SP-Ring demonstrates the poorest scalability among all parallel methods.

For the 2048px image generation task, the optimal hybrid parallel configuration becomes cfg=2, pipefusion=2, ring=2 for 8 GPUs. 
Similarly, hybrid parallel achieves a marginal speedup over a single parallel method on 4 and 8 GPUs. However, PipeFusion showcases much higher latency than SP-Ulysses and SP-Ring when using 4 or 8 GPUs, due to the additional P2P communication required by the Skip Connections between GPUs. 
This issue is mitigated when PipeFusion operates with a parallel degree of 2, highlighting its necessity for optimal performance in hybrid configurations.
As the image size increases from 1024px to 2048px, the performance gap between SP-Ring and SP-Ulysses diminishes, because of the reduced computation-to-communication ratio of the model, which allows SP-Ring to hide a larger portion of the communication overhead.

\subsubsection{Memory Efficiency of DiTs Backbone}

The maximum memory usage of different parallel approaches for the backbones of Pixart, SD3, and Flux.1 is depicted in Figure~\ref{fig:3modelmemory}. 
The figure includes memory usage for "parameters", which encompasses both the text encoder and transformers, as well as for "others", which includes activations and temporary buffers. Note that SP-Ulysses and SP-Ring exhibit similar memory consumption, denoted as SP in the figure.

Pixart, with only 0.6B transformers parameters, shows minimal variation in "parameters" memory across different parallel methods, where the memory consumption of the text encoder dominates. 
Observing the "others" memory usage reveals that DistriFusion consumes significantly more memory, particularly as image size increases. This is because DistriFusion maintains full spatial shape K, V tensors for each layer, leading to a substantial increase in memory consumption with larger images.

For SD3 and Flux, PipeFusion demonstrates the lowest memory footprint, because it only needs to store 1/pipefusion\_degree of the parameters.
This is particularly crucial for Flux.1, which has a larger number of model parameters, resulting in significantly lower memory consumption on 8$\times$GPU compared to SP. However, PipeFusion requires kv buffers for DiTs blocks belonging to its own pipeline stage, which increases the others memory usage compared to baseline and SP methods.
\textbf{\textit{The overall memory usage of PipeFusion is 32\% and 36\% of SP on 1024px and 2048px cases using Flux.1}}.


\subsubsection{Analysis}
We summarize the experimental results on DiTs backbones from two types of GPU clusters. Firstly, PipeFusion demonstrates high communication and memory efficiency as a parallel method compared to sequential parallelism and tensor parallelism. It consistently exhibits lower latency than other parallel approaches across most tasks, particularly evident when using PCIe-interconnected L40s. Additionally, it significantly reduces memory overhead on large models like Flux.1.

Secondly, adapting to different model architectures and scaling to extremely long sequences necessitates hybrid parallel. 
As shown before, no single parallel method can achieve optimal performance across different hardware and model configurations. When scaling to a 16$\times$L40 setup, only hybrid parallel further reduces latency. 
Our best hybrid practice recommendation is to prioritize cfg parallel. 
In low-bandwidth PCIe and Ethernet environments, prioritize PipeFusion, followed by SP-Ring; in high-bandwidth NVLink networks, prioritize SP-Ulysses and PipeFusion.

Thirdly, the results justify the reason for not adding tensor parallelism and Thirdly as the foundational parallel methods for xDiT.
The former exhibits poor scalability, whereas the latter incurs significant memory overhead on long sequences.

\subsection{Parallel VAE Performance Results}
\label{sec:parallvaeresults}

\begin{table*}[t]
\small
\begin{tabular}{|c|c|c|c|c|c|c|c|c|c|c|c|c|c|c|c|c|}
\hline
& \multicolumn{4}{c|}{\textbf{8$\times$L40 (40GB) 16 channels}} & \multicolumn{4}{c|}{\textbf{8$\times$L40 (40GB) 4 channels}} & \multicolumn{4}{c|}{\textbf{8$\times$A100 (80GB) 16 channels}} & \multicolumn{4}{c|}{\textbf{8$\times$A100 (80GB) 4 channels}} \\
\hline
\textbf{GPUs} & \textbf{1k} & \textbf{2k} & \textbf{4k} & \textbf{7k} & \textbf{1k} & \textbf{2k} & \textbf{4k} & \textbf{7k} & \textbf{1k} & \textbf{2k} & \textbf{4k} & \textbf{8k} & \textbf{1k} & \textbf{2k} & \textbf{4k} & \textbf{8k} \\
\hline
1 & 0.74 & 2.18 & OOM & OOM & 0.75 & 2.18 & OOM & OOM & 1.07 & 1.76 & OOM & OOM & 0.87 & 1.8 & NaN & OOM \\
\hline
2 & 0.82 & 1.95 & OOM & OOM & 1.03 & 1.99 & OOM & OOM & 2.03 & 2.71 & 11.56 & OOM & 1.95 & 2.8 & 11.75 & OOM \\
\hline
4 & 1.22 & 2.08 & 9.78 & OOM & 1.33 & 2.24 & 10.26 & OOM & 4.95 & 5.37 & 13.63 & OOM & 5.1 & 5.31 & 14.32 & OOM \\
\hline
8 & 2.08 & 3.7 & 9.66 & 68.91 & 2.28 & 2.98 & 9.94 & 72.39 & 11.75 & 12.35 & 19.64 &  146.08 & 11.45 & 11.48 & 19.67 & 149.09 \\
\hline
\end{tabular}
\caption{Elapsed Time (Sec) for Different Image Sizes, Channels, and Number of GPUs on 8$\times$L40 and 8$\times$A100.}
\label{tab:elapsed_time_channels_gpu}
\end{table*}

\begin{figure*}[htbp!]
  \centering
  \small
  \begin{minipage}{0.16\textwidth}
    \centering
  \begin{normalsize}Original\end{normalsize}\\
  1-Device\\
  FID: 22.78
  \end{minipage}
  \begin{minipage}{0.16\textwidth}
  \centering
  \begin{normalsize}PipeFusion+USP\end{normalsize}\\
  8 Devices\\
  pp=2, sp=4\\
  FID: 15.18
  \end{minipage}
  \begin{minipage}{0.16\textwidth }
  \centering
  \begin{normalsize}PipeFusion+USP\end{normalsize}\\
  8 Devices\\
  pp=4, sp=2\\
  FID: 15.96
  \end{minipage}
  \begin{minipage}{0.16\textwidth}
  \centering
  \begin{normalsize}PipeFusion\end{normalsize}\\
  2 Devices\\
  patch number=2\\
  FID: 23.60
  \end{minipage}
  \begin{minipage}{0.16\textwidth}
  \centering
  \begin{normalsize}PipeFusion\end{normalsize}\\
  4 Devices\\
  patch number=4\\
  FID: 25.23
  \end{minipage}
   \begin{minipage}{0.16\textwidth}
  \centering
  \begin{normalsize}PipeFusion\end{normalsize}\\
  8 Devices\\
  patch number=8 \\
  FID: 28.46
  \end{minipage}
  \begin{subfigure}[b]{\textwidth}
    \centering
    \includegraphics[width=0.16\textwidth]{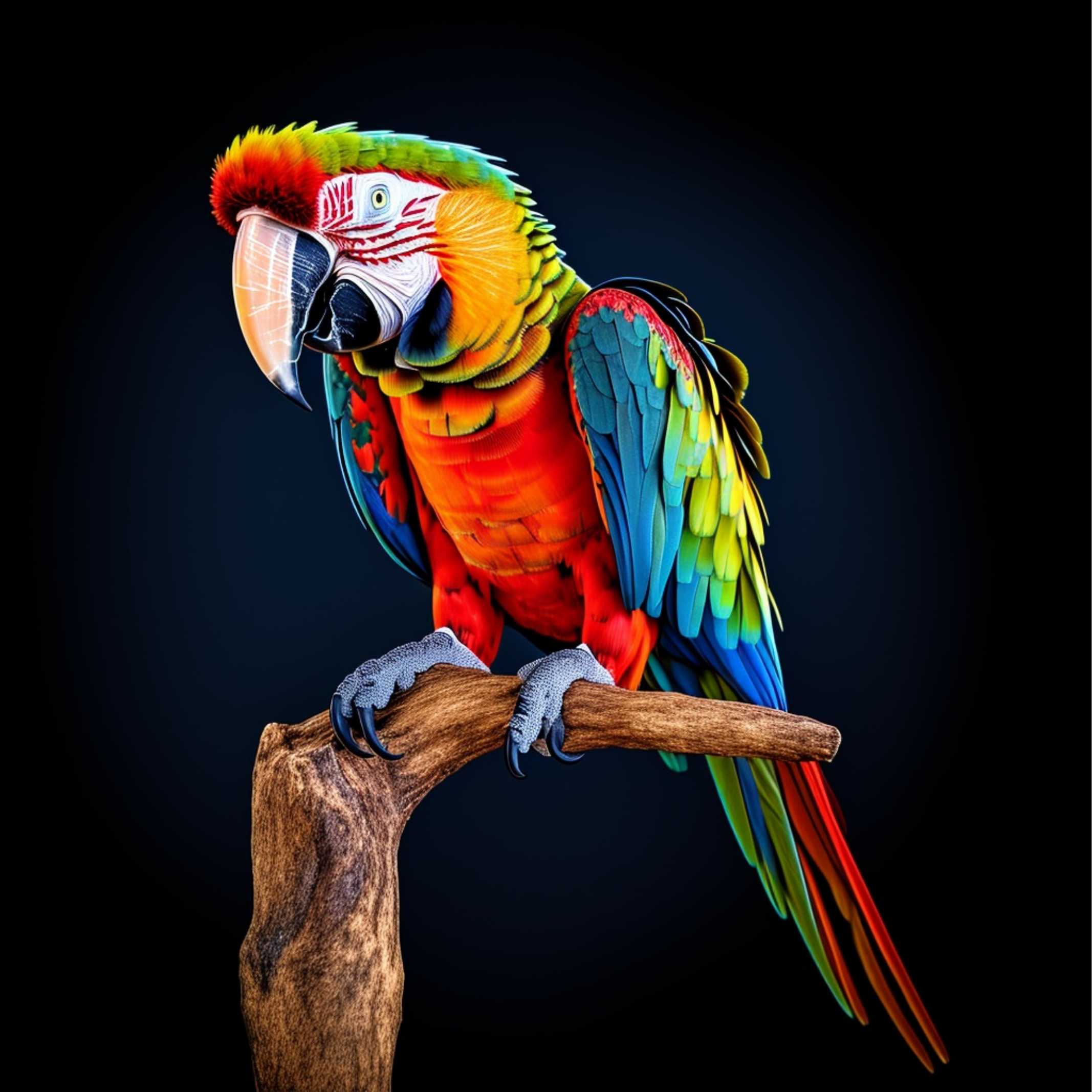}
    \includegraphics[width=0.16\textwidth]{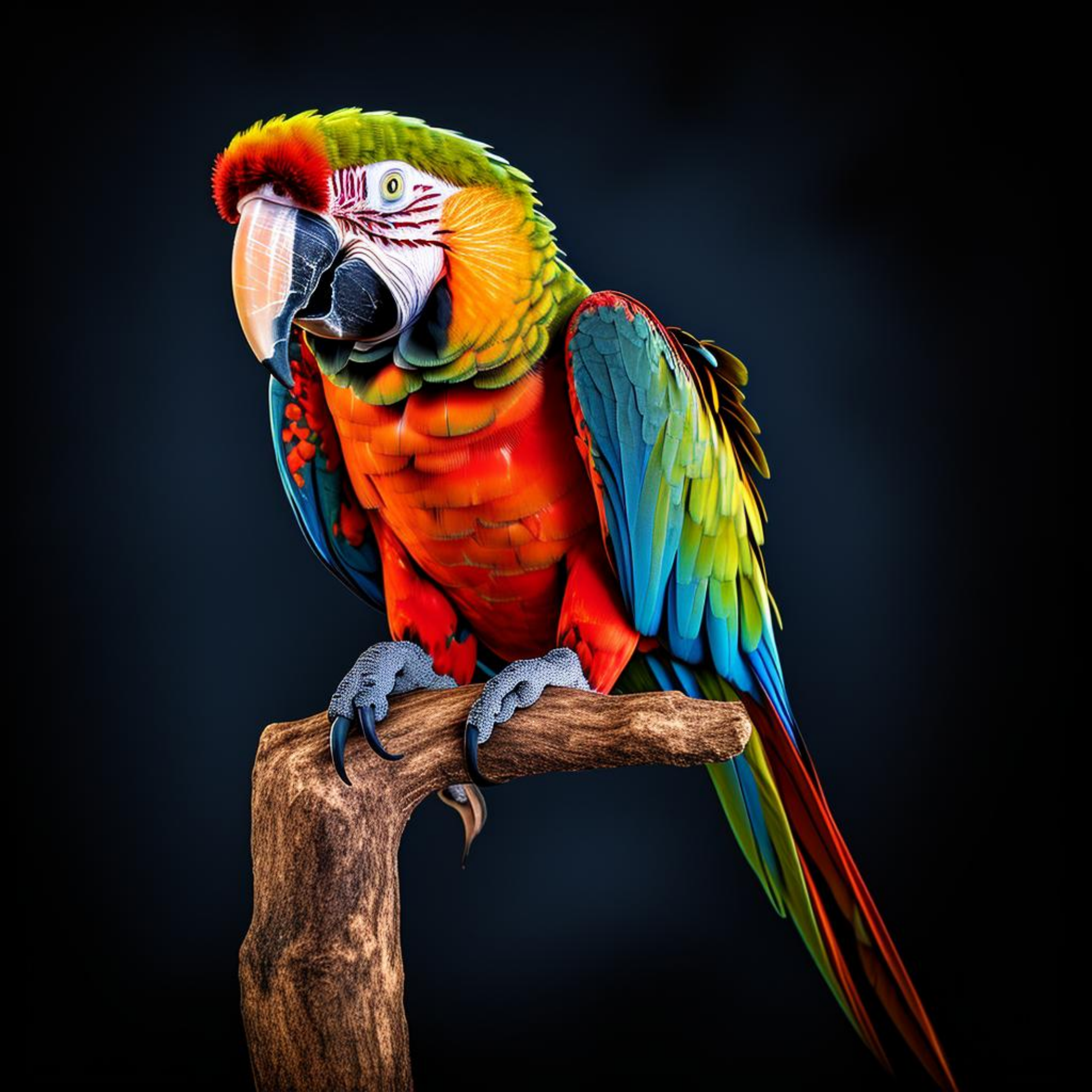}
    \includegraphics[width=0.16\textwidth]{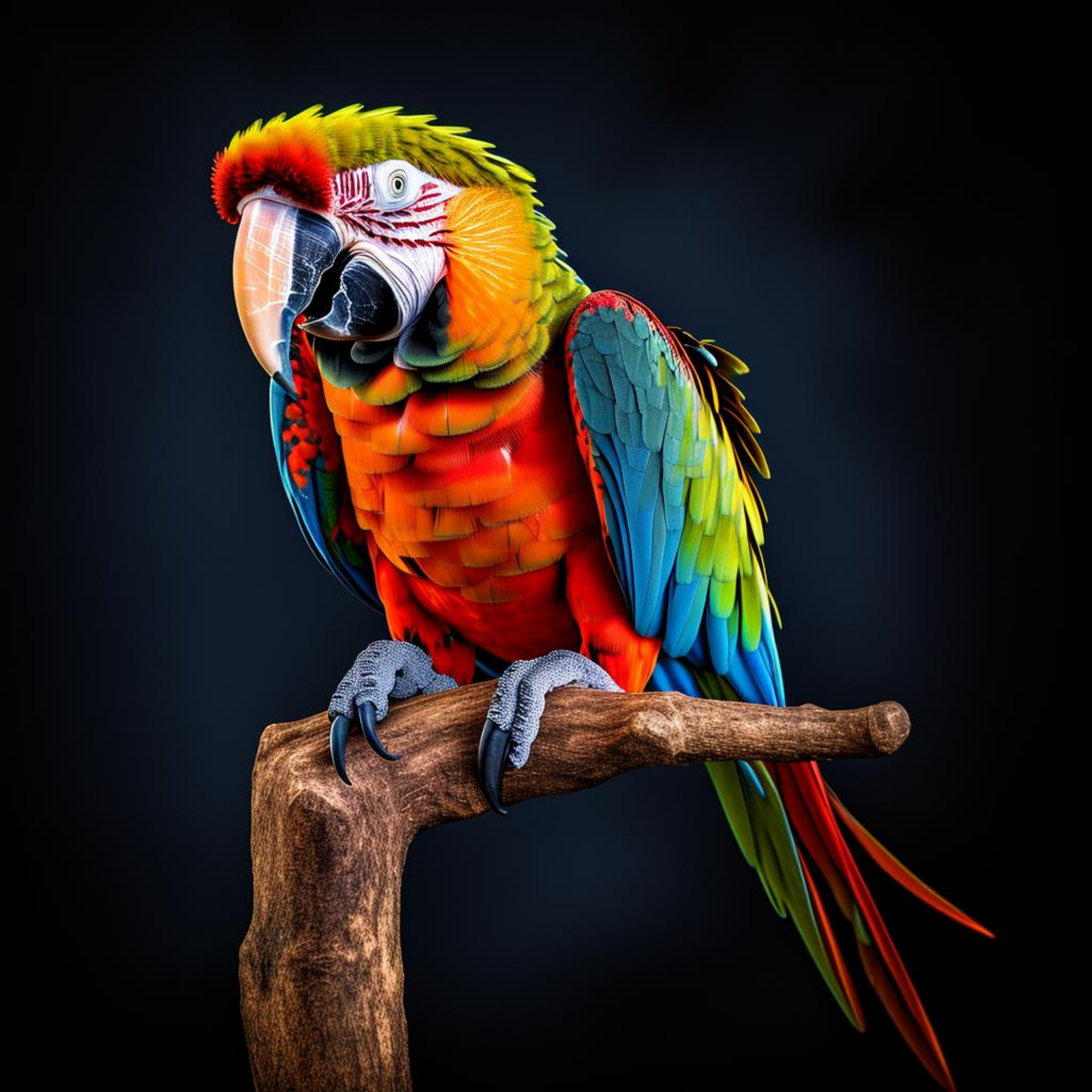}
    \includegraphics[width=0.16\textwidth]{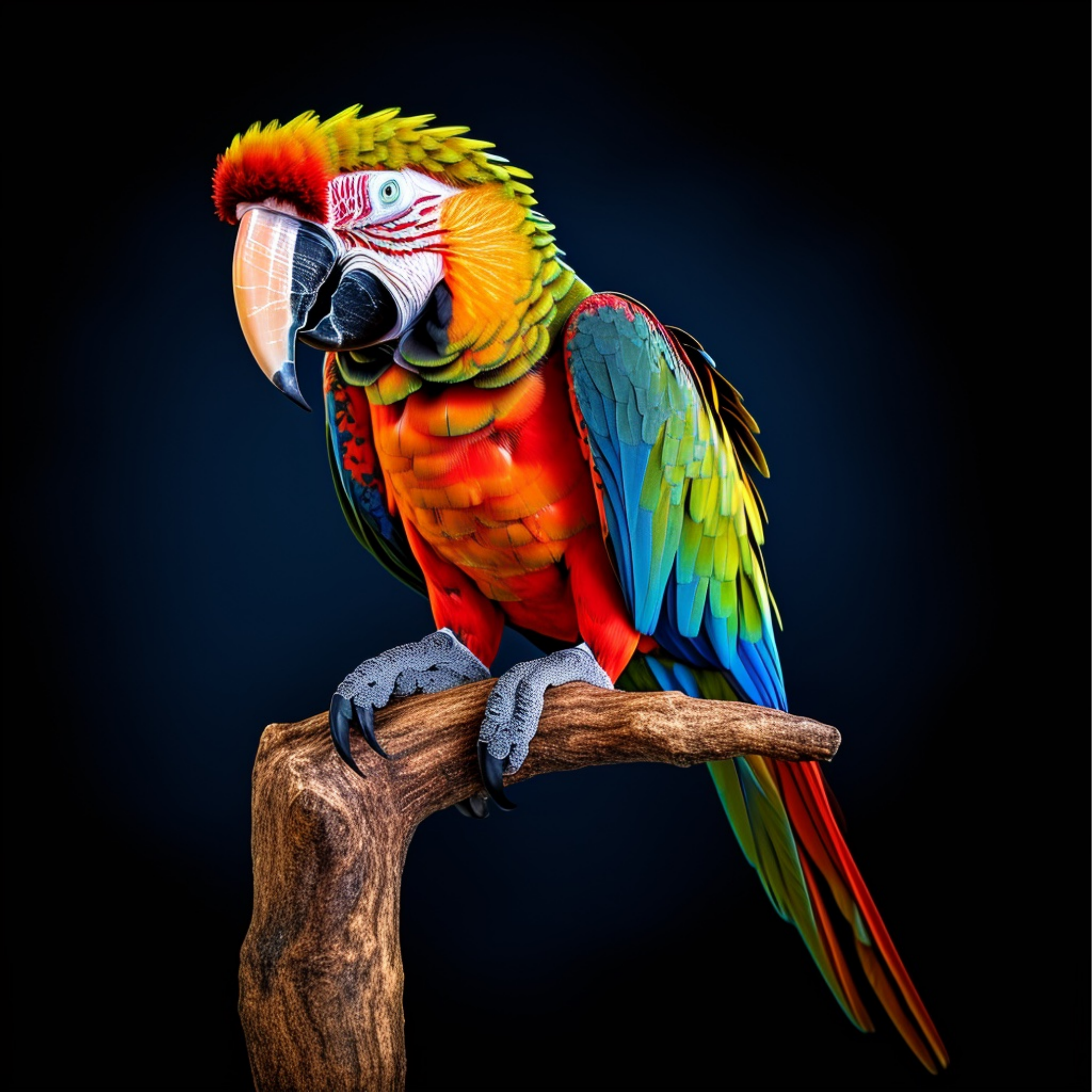}
    \includegraphics[width=0.16\textwidth]{./figures/results/parrot/pip.pdf}
    \includegraphics[width=0.16\textwidth]{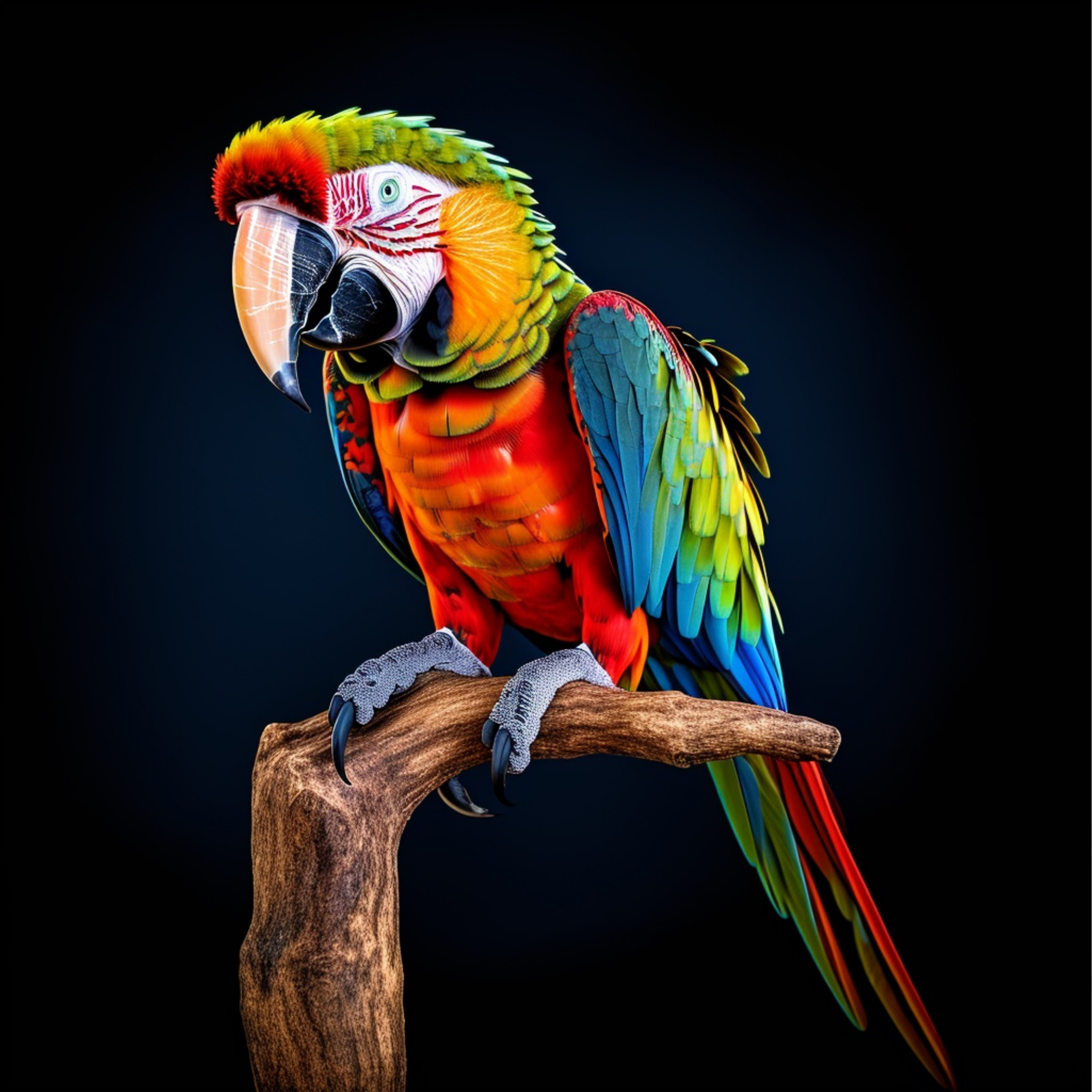}
    \parbox{0.9\textwidth}{\centering Prompt: \textit{A multi-colored parrot holding its foot up to its beak.}}
  \end{subfigure}
  \begin{subfigure}[b]{\textwidth}
    \centering
    \includegraphics[width=0.16\textwidth]{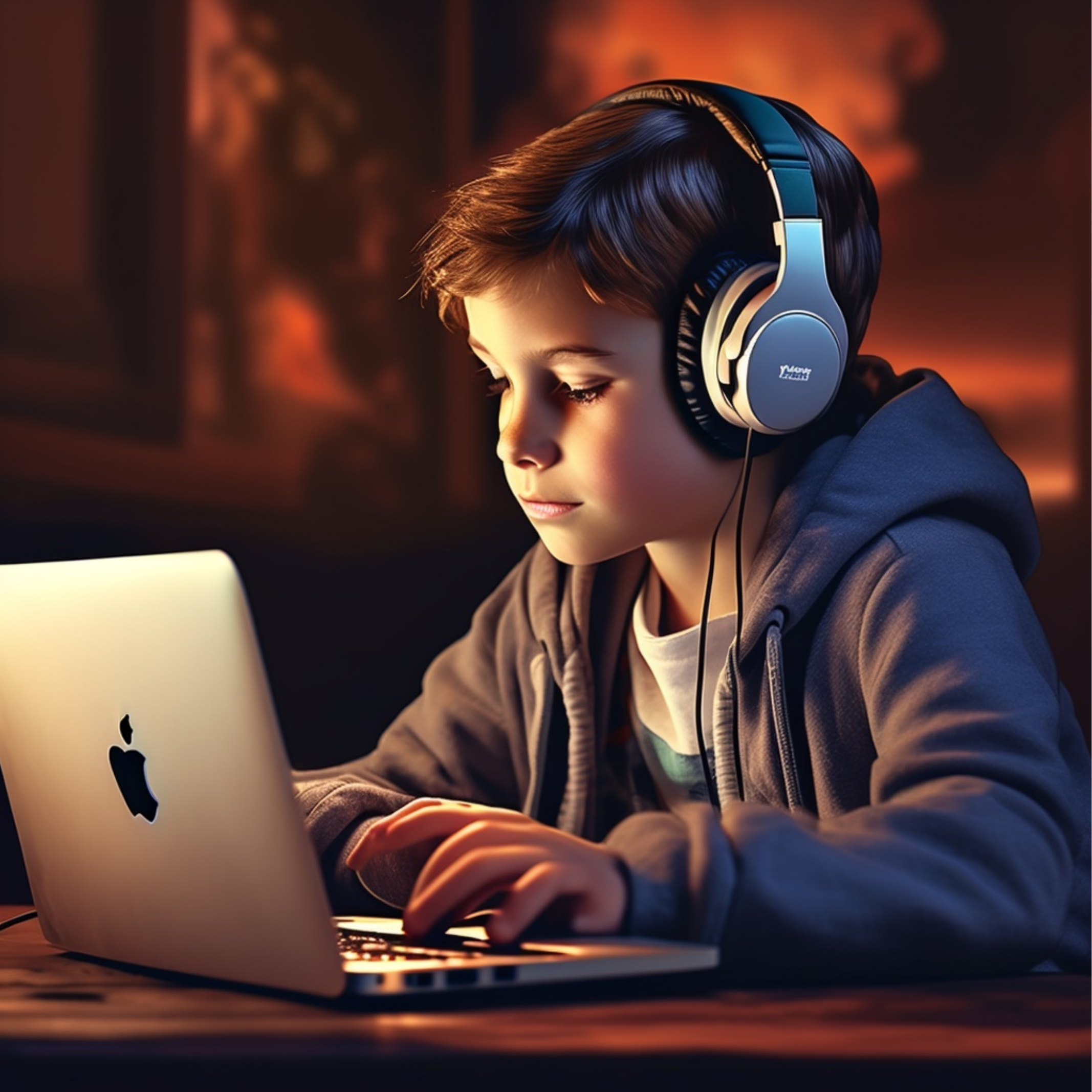}
    \includegraphics[width=0.16\textwidth]{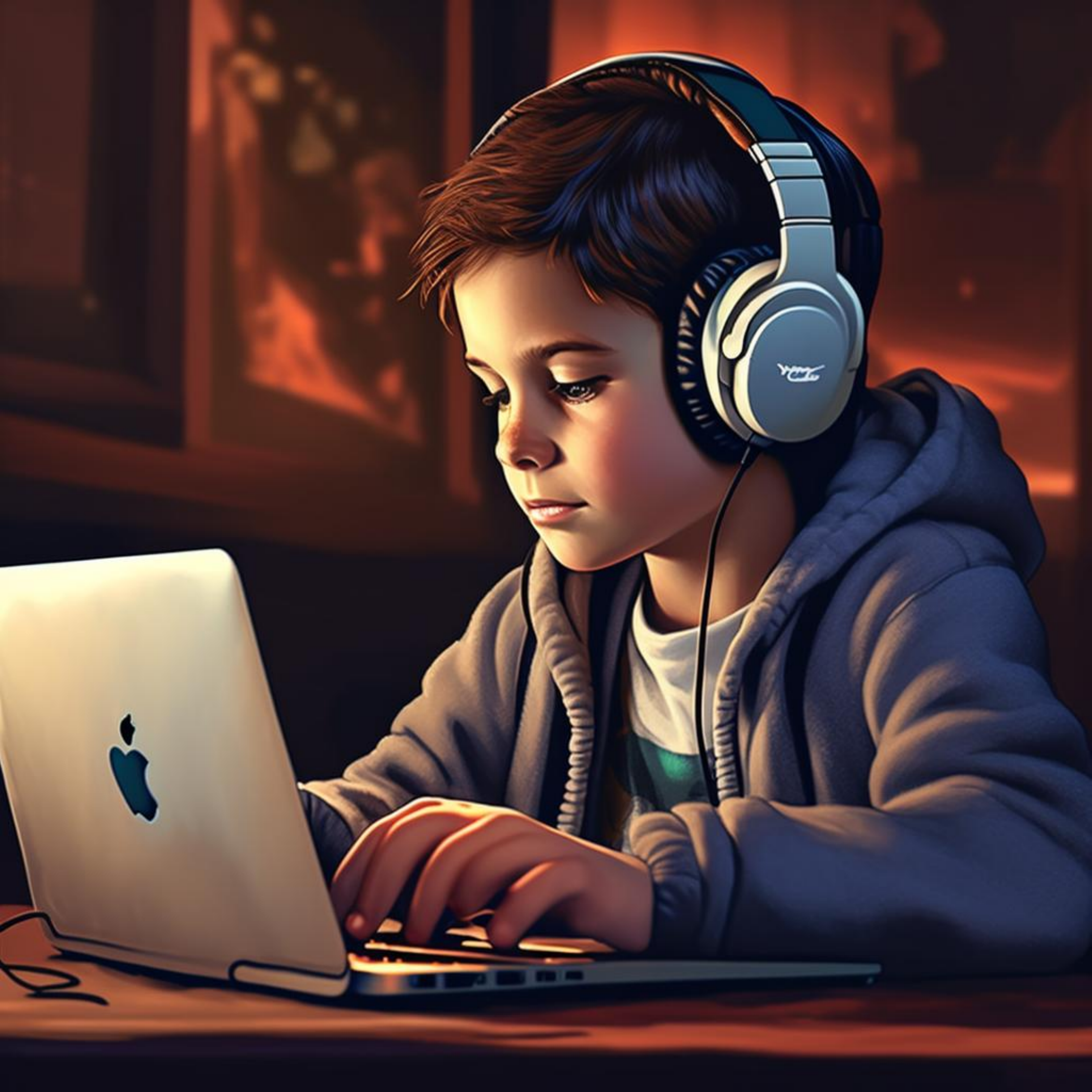}
    \includegraphics[width=0.16\textwidth]{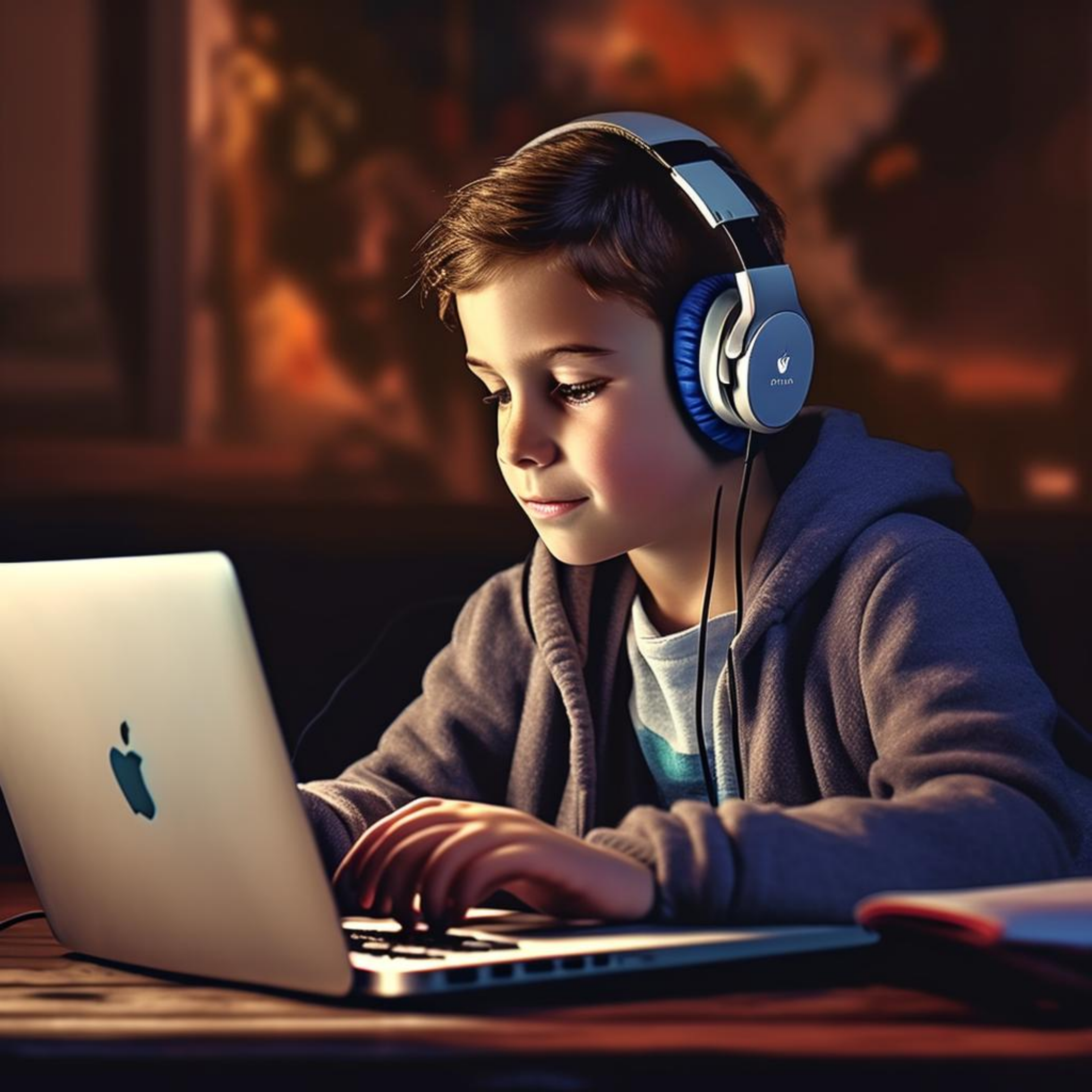}
    \includegraphics[width=0.16\textwidth]{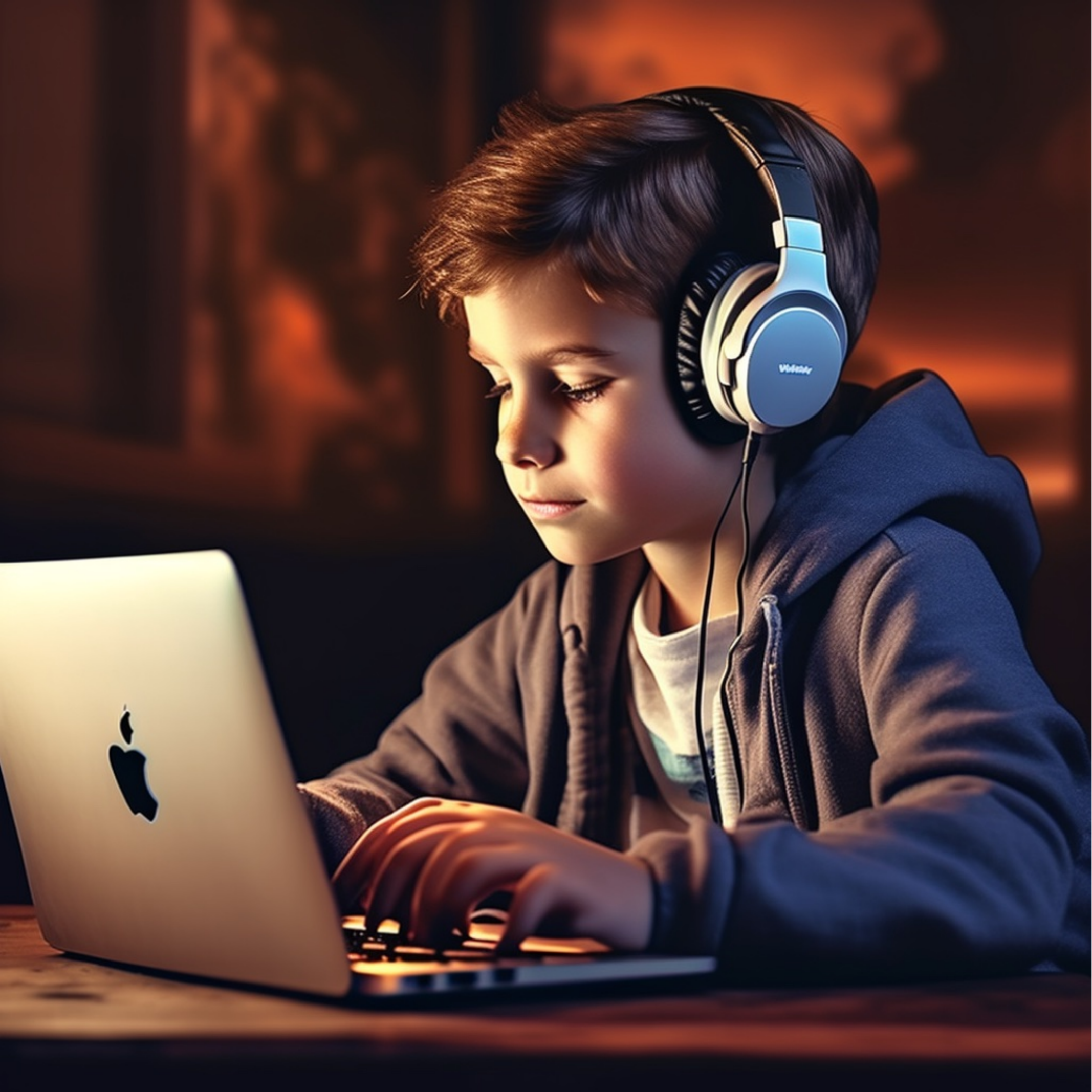}
    \includegraphics[width=0.16\textwidth]{./figures/results/kid/pip.pdf}
    \includegraphics[width=0.16\textwidth]{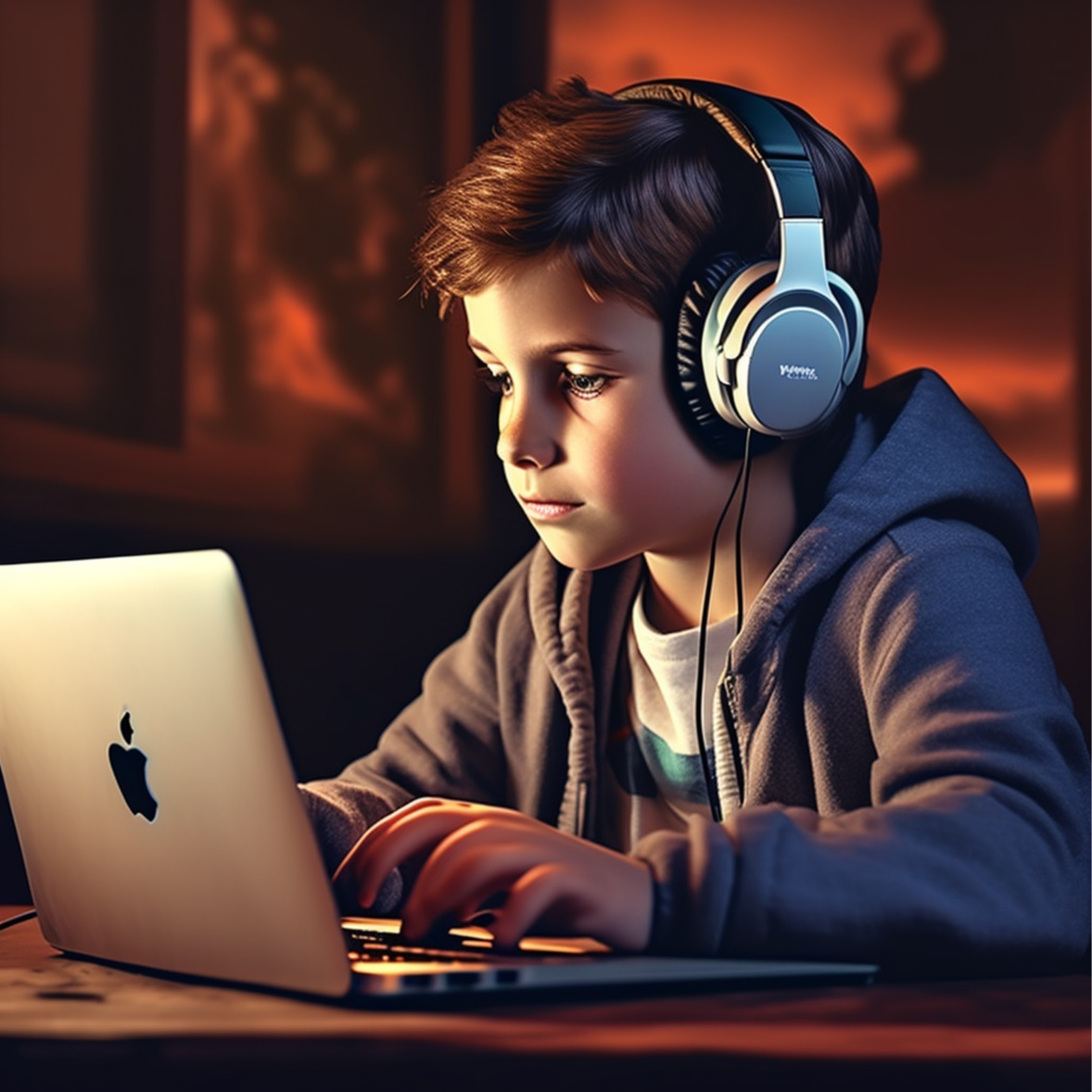}
    \parbox{0.9\textwidth}{\centering  Prompt: \textit{A kid wearing headphones and using a laptop.}}
  \end{subfigure}
  \caption{Showcases for 1024px Generation Images with hybrid parallel using Pixart. 
FID above the images is computed against the ground-truth images using Clean-FID~\cite{parmar2022aliased} 
  We use a 20-step DPM-Solver with the warmup step set to 1 for  PipeFusion and USP.
We use the COCO Captions 2014~\cite{chen2015microsoft} dataset to evaluate the FID scores. 
During the evaluation, a subset comprising 30,000 images is sampled from the validation set and resized to 256px to serve as the reference dataset. Concurrently, each experiment generates 30,000 images of 256px, each paired with a caption derived from the COCO Captions 2014 dataset, as the sample dataset. 
In PipeFusion+USP cases, "pp" and "sp" represent the parallelism of PipeFusion and USP, respectively.}
  \label{fig:quality}
\end{figure*}

The VAE, as an independent module, can be decoupled from the DiTs backbone. 
On one hand, they can be deployed on the same devices, but this may lead to memory interference. 
On the other hand, they can be deployed on different computing devices, and a output tensor of DiTs backbone is transferred to the VAE devices. The parallelism of the two modules can also be different. Therefore, this section evaluates the performance of the VAE independently.

The VAE model architecture used in the four image generation DiTs is the same.
When using different models, the shape of the input image for the VAE varies.
For the channel dimension, Pixart, SD3, and HunyuanDiT are 4, while Flux.1 is 4.
The table below shows the latency of the Parallel VAE on 8$\times$L40 and 8$\times$A100 for different generated image resolutions.

Firstly, 1-GPU baseline VAE can only support image generation at 2048px resolution on both L40 and A100.
Using 8$\times$GPU, the resolution can be increased to 7168px for L40 and 8192px for A100.
however, for a specific input, increasing parallelism does not show good scalability. 
Only at 2048px does the latency on 2$\times$L40 decrease compared to 1$\times$L40; in all other parallel cases, the latency is higher than on 1 GPU. 
This is because the computation of the convolution operators is relatively small, and the communication cost introduced by patch parallel is relatively high. 
Therefore, Parallel VAE should be applied to alleviate OOM issue and make the model successfully finish image decoding, rather than to accelerate the execution speed.

\subsection{Quality Results}


We investigate the generation accuracy of different parallel methods, especially hybrid parallel involving PipeFusion and USP. In Figure~\ref{fig:quality}, we present qualitative visual results for 1024px image generation using Pixart-alpha. For all the experiments, only 1 warmup iteration was employed. 
The images generated with xDiT are virtually indistinguishable from the original images to the human eye, across various settings. Furthermore, we include the Fréchet Inception Distance (FID)~\cite{heusel2017gans} in the figure (a lower FID score is preferable). 
This also validates the correctness of our hybrid parallel approach in Sec.~\ref{sec:hybrid_paralll}.

Regarding the FID metric, it is evident that applying parallelism may not affect the FID scores in comparison to the baseline.
Notably, the hybrid parallel approach (PipeFusion+USP) even demonstrates superior performance over the baseline in FID. 
Many studies\cite{betzalel2022study}\cite{chen2023pixart}\cite{kirstain2023pick}\cite{podell2023sdxl} suggest that the FID score may not accurately reflect the visual quality of generated images. 
Consequently, we propose that the accuracy of parallel approaches should be assessed based on the observed visual quality of the generated images, rather than being solely dependent on the FID metric.

\section{Conclusion}
This paper introduces xDiT, a DiTs parallel inference engine, which encompasses several key innovations. Firstly, we adapt Sequence Parallel (SP) to various state-of-the-art (SOTA) DiTs architectures. Secondly, we review PipeFusion, a Patch-level Pipeline Parallel, and explore its application across different DiTs architectures. Thirdly, based on an extensive investigation of existing parallel strategies, we pick PipeFusion, SP, and CFG parallel as foundational parallel approaches and design a method to flexibly combine them in a hybrid parallel manner. Fourthly, we implement patch parallelism for the VAE module to prevent out-of-memory (OOM) issues.

We evaluate xDiT on four image generation and one video generation DiTs models using 16$\times$L40 and 8$\times$A100 GPU configurations. Our results confirm the communication and memory advantages of PipeFusion over other methods and demonstrate that xDiT can efficiently scale DiTs inference to large-scale clusters.

\bibliographystyle{plain}
\bibliography{references}

\end{document}